\documentclass[10pt,aps,prd,twocolumn,notitlepage,bibnotes,longbibliography,
floatfix,showpacs,preprintnumbers,citeautoscript,superscriptaddress]{revtex4-1}

\usepackage[english]{babel}		
\makeatletter\AtBeginDocument{\let\@elt\relax}\makeatother 
\usepackage{amsmath}		
\usepackage{amssymb}		
\usepackage{mathtools}		
\usepackage{bbm}			

\usepackage{array}			
\usepackage{graphicx}		
\usepackage{graphics}		
\DeclareGraphicsExtensions{.jpg,.jpeg,.png,.pdf,.eps}	

\usepackage[pdfstartview=FitV, pdfusetitle,
			bookmarks=true,bookmarksnumbered=false,bookmarksopen=false,
			breaklinks=true,pdfborder={0 0 0},backref=false,colorlinks=true, 
			linkcolor=blue, citecolor=blue, urlcolor=blue]{hyperref}

\renewcommand{\theequation}{\arabic{equation}}

\makeatletter
\newcommand*{\Equation}{\@ifstar\sEquation\oEquation}
\newcommand{\sEquation}[1]{\begin{equation*}#1\end{equation*}}
\newcommand{\oEquation}[2]{  \begin{equation}\label{#1}#2\end{equation} }
\makeatother
\newcommand{\Align}[2]{\begin{align}\label{#1}#2\end{align}}
\newcommand{\SubAlign}[2]{\begin{subequations}\label{#1}
						  \begin{align}#2\end{align}\end{subequations}}

\newcommand{\bs}{\boldsymbol}

\newcommand{\Figref}[1]{Fig.~\ref{#1}}

\newcommand{\Eqref}[1]{\eqref{#1}}

\newcommand{\eg}{{\it e.g.~}}

\newcommand{\ie}{{\it i.e.~}}

\newcommand{\Identity}{\mathbbm{1}}

\newcommand{\groupO}[1]{\mathrm{O}(#1)}   			
\newcommand{\groupU}[1]{\mathrm{U}(#1)}  			
\newcommand{\groupSU}[1]{\mathrm{SU}(#1)}			
\newcommand{\groupZ}[1]{\mathbb{Z}_{#1}} 			
\newcommand{\groupR}[1]{\mathbb{R}^{#1}} 			
\newcommand{\groupS}[1]{\mathbb{S}^{#1}} 			
\newcommand{\groupUZ}{\groupU{1}\!\times\!\groupZ{2}} 			
\newcommand{\groupUZthree}{\groupU{1}\!\times\!\groupZ{3}} 			
\newcommand{\groupUZn}{\groupU{1}\!\times\!\groupZ{n}} 			
\newcommand{\groupUU}{\groupU{1}\!\times\!\groupU{1}} 			

\renewcommand\Im{\mathrm{Im}}
\newcommand{\tr}{\mathrm{tr}\,}

\newcommand{\x}{\bs x}
\newcommand{\Grad}{{\bs\nabla}}

\newcommand{\Curl}{{\bs\nabla}\times}
\newcommand{\Norm}[1]{{\lVert #1 \rVert}}
\newcommand{\ScalarProd}[2]{\left\langle #1,#2\right\rangle}

\newcommand{\F}{\mathcal{F}}
\newcommand{\Q}{\mathcal{Q}}
\newcommand{\Tc}[1]{T_c^{#1}}   			

\newcommand{\sis}{{s\!+\!is}}

\newcommand{\sid}{{s\!+\!id}}

\newcommand{\dig}{{d\!+\!ig}}
\newcommand{\pip}{{p\!+\!ip}}

\newcommand{\D}{{\bs D}}

\newcommand{\A}{{\bs A}}

\newcommand{\B}{{\bs B}}

\newcommand{\J}{{\bs J}}
\renewcommand{\j}{{\bs j}}

\newcommand{\cR}{\mathcal{R}}
\newcommand{\hk}{\hat{k}}
\newcommand{\vn}{{\bs n}}
\newcommand{\m}{{\bs m}}
\newcommand{\n}{{\bs n}}

\begin{document}
\graphicspath{{../Final-plots/}}
\epstopdfsetup{outdir=./}
\title{Effective model and Magnetic Properties of the Resistive Electron Quadrupling State}

\author{Julien Garaud}
\email{garaud.phys@gmail.com}
\affiliation{Institut Denis Poisson CNRS-UMR 7013,  
			 Universit\'e de Tours, 37200 Tours, France}
\affiliation{Nordita, Stockholm University, Roslagstullsbacken 23, 
			 SE-106 91 Stockholm, Sweden}

\author{Egor Babaev}
\email{babaev.egor@gmail.com}
\affiliation{Department of Physics, KTH Royal Institute of Technology, 
			 SE-106 91 Stockholm, Sweden}

\date{\today}
\begin{abstract}

Recent experiments [V.~Grinenko {\it et al.} 
\href{http://doi.org/10.1038/s41567-021-01350-9}{Nat. Phys. {\bf 17}, 1254 (2021)}] 
reported the observation of a condensate of four-fermion composites. This is a resistive  
state that spontaneously breaks the time-reversal symmetry, leading to unconventional 
magnetic properties, detected in muon spin rotation experiments and by the appearance of 
a spontaneous Nernst effect. 
In this work, we derive an effective model for the four-fermion order parameter that 
describes the observed spontaneous magnetic fields in this state. We show that this model, 
which is alike to the Faddeev-Skyrme model can host skyrmions: magnetic-flux-carrying 
topological excitations.

\end{abstract}

\maketitle

Recent experiments \cite{Grinenko.Weston.ea:21} reported the observation of a fermion 
quadrupling state in the multiband material: hole-doped Ba$_{1-x}$K$_x$Fe$_2$As$_2$. 
This resistive state, coined quartic bosonic metal, is a condensate with an anticorrelated 
flow of pairs of Cooper pairs belonging to different bands. In contrast to superconductors, 
which break the $\groupU{1}$ gauge symmetry, this state spontaneously breaks the two-fold 
($\groupZ{2}$) time-reversal symmetry. This raises the question of the properties of such 
states.

An effective model can describe the properties of condensates at large length scales. 
For a pair condensate, the effective model is the celebrated Ginzburg-Landau theory 
which has been extensively studied since the second half of the last century. The question 
of effective models describing the fermion quadruplet quartic metal is more subtle. 
In this paper, we derive an effective long-wavelength model for the resistive quartic 
state reported in Ba$_{1-x}$K$_x$Fe$_2$As$_2$. Based on this, we report the key properties 
of that state: Namely its magnetic properties and the nature of the topological excitations 
it supports.

At low temperatures, the compound is a superconductor characterized by Cooper pair condensates 
$\Delta_a$, forming in the different bands labeled by $a$. Importantly this superconductor 
breaks the time-reversal symmetry \cite{Grinenko.Materne.ea:17,Grinenko.Sarkar.ea:20}, so that 
the total symmetry broken by the low-temperature state is $\groupUZ$. The analysis of the 
magnitude and polarization of spontaneous magnetic fields \cite{Grinenko.Sarkar.ea:20,
Vadimov.Silaev:18,Speight.Winyard.ea:21} indicates a spin-singlet superconducting state that 
breaks the time-reversal symmetry. It is the so-called $\sis$ state which has two energetically 
equivalent locking of the relative phase $\theta_b-\theta_a$ between the superconducting 
gaps in different components $\Delta_{a,b}$.

The mechanism responsible for the appearance of the quartic metal is the following: 
The standard assumption of the Bardeen-Cooper-Schrieffer theory is a mean-field 
approximation for the fields quadratic in fermions:  This assumption eliminates, 
by construction, the possibility for fermion quadrupling. The resulting theory yields 
the phase diagram of such a superconductor, which is typically a dome of the $\sis$ state 
between two different superconducting states \cite{Stanev.Tesanovic:10,
Carlstrom.Garaud.ea:11a,Maiti.Chubukov:13,Silaev.Garaud.ea:17,Boeker.Volkov.ea:17}.
It was pointed out in \cite{Babaev.Sudbo.ea:04,Babaev:04}, that relaxing the mean-field 
approximation in a multicomponent fermion pairing theory results in a phase diagram with the 
appearance of fermion quadrupling condensates. The large-scale Monte Carlo calculations of 
$\groupUZ$ states demonstrated that the discrete $\groupZ{2}$ transition can exceed the
superconducting $\groupU{1}$ transition: $\Tc{}<\Tc{\groupZ{2}}$ \cite{Bojesen.Babaev.ea:13,
Bojesen.Babaev.ea:14,Carlstrom.Babaev:15}.

The spontaneous breakdown of the time-reversal symmetry in the resistive state  of 
Ba$_{1-x}$K$_x$Fe$_2$As$_2$, at the doping level $x\approx0.8$ \cite{Grinenko.Weston.ea:21} 
dictates that the averages of the pairing order parameters $\Delta_a$ are zero, but that 
there exists a nonzero order parameter which is fourth order in the fermionic fields. 
The quadrupling order parameter is proportional to the product of pairing order parameters 
in different bands $\Delta_a^*\Delta_b$ .
Such an order parameter implies an anticorrelation in the flows of the components $a$ 
and $b$. Crucially, although these types of counterflows do not represent superconductivity, 
they are generally coupled to the magnetic field when the densities of the counterflowing 
charged components are unequal. An effective model should account for this coupling, 
and should be different from the Ginzburg-Landau model of a Meissner state.

Below we derive such an \emph{effective} theory, based on the mean-field approximation 
for the four-fermion order parameter.
We demonstrate that, in an inhomogeneous sample, the model supports spontaneous magnetic 
fields, consistently with the experimental results \cite{Grinenko.Weston.ea:21}. 
It also predicts the existence of topological excitations 
carrying a quantized magnetic flux, in the form of skyrmions.

We derive our effective model for a state with composite order, from a generic model of a 
superconductor with a two-component order parameter $\Psi$, with 
$\Psi^\dagger:=(\psi_1^*,\psi_2^*)$. The detailed derivation from the microscopic theory can 
be found in Supplemental Material \cite{*[{}] [{}] Supplementary-arxiv}. 
The generic Ginzburg-Landau free-energy density for a two-component superconductor reads as
\Equation{Eq:FreeEnergy:0}{
\F(\Psi,\A)= \frac{\B^2}{2}
+\frac{k_{ab,ij}}{2} (D_i\psi_a )^*D_j \psi_b  
+ V(\Psi^\dagger,\Psi)	\,,
}
where $V(\Psi^\dagger,\Psi)$ is the potential energy term. The repeated indices are 
implicitly summed over, and the indices $i,j$ denote the spatial coordinates while 
$a,b$ label the different components. The individual condensates are coupled to the 
vector potential $\A$, of the magnetic field $\B=\Curl\A$, via the gauge derivative 
$\D=\Grad+ie\A$ in the kinetic term.
In this work, we focus on two-component models that break multiple symmetries.
The symmetry breaking is encoded in the potential term $V(\Psi^\dagger,\Psi)$ 
which explicitly reduces the global $\groupSU{2}$ symmetry of a doublet of complex 
order parameters down to a smaller symmetry group.
For example, the $\groupSU{2}$ symmetry is broken down to $\groupUZ$, for a superconductor 
that breaks time-reversal symmetry such as $\sis$, $\sid$, $\dig$, $\pip$, or down to 
$\groupUZthree$ symmetry as was suggested for some nematic superconductors 
\cite{Cho.Shen.ea:20}.
The composite order of interest arises if the fluctuations-driven restoration 
of the local gauge symmetry occurs without restoring the other broken symmetries.
The existence of a composite order was demonstrated in systems featuring $\groupUU$ 
\cite{Babaev.Sudbo.ea:04,Babaev:04} and $\groupSU{2}$ \cite{Kuklov.Matsumoto.ea:08,
Herland.Bojesen.ea:13} symmetries and from these calculations it follows that composite 
order also exists for $\groupUZn$ symmetries.
While most of our results qualitatively apply to all of the above mentioned pairing 
mechanisms, we focus below on the case of the broken time-reversal symmetry $\groupUZ$, 
and in particular on the $\sis$ state, motivated by the  experiment on 
Ba$_{1-x}$K$_x$Fe$_2$As$_2$ \cite{Grinenko.Weston.ea:21}.
Other related states with composite order were discussed in 
\cite{Agterberg.Tsunetsugu:08,Berg.Fradkin.ea:09,Kuklov.Prokofev.ea:06,Erten.Chang.ea:17,
Fleurov.Kuklov:19,Shaffer.Wang.ea:21,Buessen.Sorn.ea:21,Fernandes.Fu:21,Chung.Kim:22,
DrouinTouchette.Orth.ea:22}.

At the microscopic level, the minimal model features three distinct superconducting 
gaps $\Delta_{1,2,3}$ in three different bands, and the pairing that leads to the 
time-reversal symmetry breaking states is dominated by the competition between different 
interband repulsion channels \cite{Stanev.Tesanovic:10,Maiti.Chubukov:13,
Boeker.Volkov.ea:17}. In the case of an interband-dominated repulsive pairing, only two 
fields $\psi_{1,2}$ appear in the effective Ginzburg-Landau model for the superconducting 
state, see \eg \cite{Maiti.Chubukov:13,Garaud.Silaev.ea:16,Garaud.Silaev.ea:17}. 
When starting from the microscopic three-band model, the relevant two-component 
Ginzburg-Landau theory features mixed-gradient terms, which can be eliminated by a 
linear transformation to new fields see \eg \cite{Garaud.Silaev.ea:17,
Garaud.Corticelli.ea:18a}, and the Supplemental Material \cite{Supplementary-arxiv}.
The resulting Ginzburg-Landau theory is characterized by the free-energy $F/F_0=\int\F$ 
whose density reads as
\Equation{Eq:FreeEnergy}{
\F(\Psi,\A)= \frac{\B^2}{2}
+\frac{1}{2}|\D\Psi|^2 
+ V(\Psi^\dagger,\Psi)	\,.
}

To account for the four-fermion state, the Ginzburg-Landau theory \Eqref{Eq:FreeEnergy} 
is first mapped onto a model that couples the supercurrent 
$\J=e\Im\big(\Psi^\dagger\D\Psi\big)$ to a real 3-vector $\m$. It is defined as the 
projection of the superconducting degrees of freedom $\Psi$ onto spin-1/2 Pauli matrices 
${\bs \sigma}$: $\m =\Psi^\dagger{\bs \sigma}\Psi$; hence this is an order parameter 
which is fourth order in the fermionic fields. This order parameter depends on the relative 
phase between the original complex fields, and does not depend on the superconducting degree 
of freedom: the phase sum. The norm of $\m$ is related to the total 
density squared $\Norm{\m}\equiv\varrho^2=\Psi^\dagger\Psi$.   
In terms of $\J$ and $\m$, the free energy reads as 
\cite{[{}][{}] Supplementary-arxiv} 
\Align{Eq:M:FreeEnergy:1}{
&\F= \frac{1}{2}\left[ \epsilon_{kij}\left\lbrace 
\nabla_i\left(\frac{\J_j}{e^2\varrho^2}\right) -\frac{1}{4e\varrho^6}
	\m\cdot\partial_i\m\times\partial_j\m \right\rbrace\right]^2\nonumber\\
& + \frac{\J^2}{2e^2\varrho^2} + \frac{1}{8\varrho^2}\big(\Grad\m\big)^2 + V(\m)	\,,
}
where $\epsilon$ is the rank-3 Levi-Civita symbol. 
The term in the square brackets in \Eqref{Eq:M:FreeEnergy:1} is the magnetic field 
expressed through gradients of the matter fields. The first term there, is the 
contribution of the Meissner current $\J$ to the magnetic field, while the second 
term accounts for the interband counterflow \cite{Babaev.Faddeev.ea:02,
Garaud.Carlstrom.ea:13}:
\Equation{Eq:M:Magnetic}{
\B = \Curl\!\left(\frac{\J}{e^2\varrho^2}\right) 
	-\frac{\epsilon_{abc}}{4e\varrho^6}
	m_a\Grad m_b\times\Grad m_c	\,.
}
The second term is particularly important: It is related to the counterflow of two 
components, since it has a form of gradients of the composite field $\psi_a^*\psi_b$, 
\ie it depends on gradients of the relative phase between components.
A counterflow of two identical charged components results in no charge transfer and 
hence does not couple to the magnetic field. However, if the densities of the components 
are locally imbalanced, the charge transport occurs. Thus the coupling to the magnetic 
field involves a dependence of the relative density gradients.

Next, the low-temperature model \Eqref{Eq:M:FreeEnergy:1}, which microscopic derivation 
is given in the Supplemental Material \cite{Supplementary-arxiv}, is used to obtain an 
effective model of the fermion quadrupling phase. The fermion quadrupling phase identified 
in \cite{Grinenko.Weston.ea:21,Bojesen.Babaev.ea:13,Bojesen.Babaev.ea:14} is resistive. 
This is caused by the disorder of the superconducting phase due to the proliferation of 
topological defects. 
The effective model of the resulting fermion quadrupling state is obtained  by removing 
the superconducting degrees of freedom from \Eqref{Eq:M:FreeEnergy:1}. Indeed, as 
demonstrated in Monte Carlo calculations, their prefactors are renormalized to zero  
\cite{Smorgrav.Babaev.ea:05,Smiseth.Smorgrav.ea:05,Kuklov.Matsumoto.ea:08,
Kuklov.Prokofev.ea:06,Smiseth.Smorgrav.ea:05,Bojesen.Babaev.ea:13,Bojesen.Babaev.ea:14,
Herland.Bojesen.ea:13,Weston.Babaev:21,Grinenko.Weston.ea:21}. It follows that the Meissner 
current vanishes ($\J=0$), while the currents associated with gradients of the fermion 
quadrupling order parameter $\m$ do not. 
Assuming that the critical temperatures of the $\groupZ{2}$ and $\groupU{1}$ transitions 
are well separated, the free energy of the fermion quadrupling state can be written as 
\SubAlign{Eq:M:FreeEnergy:2}{
\F(\m)&=  \frac{\big(\m\cdot\partial_i\m\times\partial_j\m  \big)^2}{16e^2\Norm{\m}^6}
  + \frac{\big(\Grad\m\big)^2}{8\Norm{\m}}	+ V(\m)	\,, \\
\text{where}&~~V(\m)= \sum_{\mathclap{a=0,x,y,z}}\alpha^\m_a m_a + \frac{1}{2}~
\sum_{\mathclap{a,b=0,x,y,z}}\beta^\m_{ab} m_am_b \,. \label{Eq:M:Potential}
}
Here the component $m_0$ stands for the magnitude of $\m$, $m_0 := \Norm{\m}$ 
(see details of the microscopic expressions for the coefficients in 
\cite{Supplementary-arxiv}).
%
The first term in \Eqref{Eq:M:FreeEnergy:2} has to be retained because it depends only 
on the relative phases and densities of the original superconducting fields. 
Hence it cannot vanish at superconducting phase transition, when $\Tc{}<\Tc{\groupZ{2}}$ 
\footnote{
This is because the topological charge of composite, single-quantum superconducting vortex 
has a winding only in the phase sum. Hence it cannot restore order in the fields that 
depend only on the relative phases and relative densities.
}.
The fermion quadrupling phase reported in \cite{Grinenko.Weston.ea:21} breaks the 
time-reversal symmetry. Hence the potential term \Eqref{Eq:M:Potential} breaks the 
symmetry associated with the vector $\m$ down to $\groupZ{2}$. 
In the original Ginzburg-Landau model \Eqref{Eq:FreeEnergy}, the time-reversal operation 
is the complex conjugation of the superconducting condensates $\Psi$. Correspondingly, 
for the soft modulus vector it is a reflection of $\m$ on the $xz$ plane of the target 
space: 
\Equation{Eq:TRS}{
{\cal T}(\Psi)=\Psi^* ~~\Leftrightarrow~~
{\cal T}(\m)=(m_x,-m_y,m_z)	\,.
}
This means that the states that break the time-reversal symmetry must have $m_y\neq0$. 
This is, for example, enforced by $\beta^\m_{xx}>0$, since it penalizes $m_x^2$.
The other details of the analysis of the potential can be found in the Supplemental 
Material \cite{Supplementary-arxiv}. The essential features can be qualitatively summarized 
as follows: First, all of the coefficients involving a $y$ index vanish: 
$\alpha^\m_y=\beta^\m_{ay}=\beta^\m_{ya}=0$. Moreover, the criterion for the condensation 
is $\alpha^{\m\,2}_0<\alpha^{\m\,2}_x+\alpha^{\m\,2}_z$, and also 
$\beta^\m_{00},\beta^\m_{zz}>0$.

The quadrupling phase appears when the mean-field approximation for the pairing fields is 
relaxed. The model \Eqref{Eq:M:FreeEnergy:2} can be viewed as a mean-field approximation 
for the fermion quadrupling fields in a resistive state; such as the $\groupZ{2}$-metal 
reported in \cite{Grinenko.Weston.ea:21}.
Since superconducting currents are absent in the resistive state, the magnetic field caused 
by the gradients in the fermion quadrupling fields becomes 
\Equation{Eq:M:Magnetic:1}{
\B = -\frac{\epsilon_{abc}
	m_a\Grad m_b\times\Grad m_c  }{4e\Norm{\m}^3}	\,.
}
In two spatial dimensions, the topological invariant, which is associated with 
the degree of the maps $\m/\Norm{m}: \groupS{2}\mapsto \groupS{2}_\m$, reads as   
\Equation{Eq:M:Charge}{
   \Q(\m)=\frac{1}{4\pi} \int_{\groupR{2}} \frac{
   \m\cdot\partial_x \m\times \partial_y \m}{\Norm{\m}^3}\,\,
  d x d y \,.
}
The integrand is obviously ill defined when $\Norm{\m}=0$. However, whenever 
$\Norm{\m}\neq0$, the corresponding configuration has an integer topological 
charge $\Q(\m)\in\groupZ{}$; this suggests that the model can host skyrmion topological 
excitations. 
Note that in three dimensions the model is characterized by another invariant, 
the Hopf invariant, which is associated with the maps $\groupS{3}\mapsto \groupS{2}_\m$. 
This suggests the existence of hopfions, but it is beyond the scope of the current 
discussion.

\begin{figure}[!tb]
\hbox to \linewidth{ \hss
\includegraphics[width=0.95\linewidth]{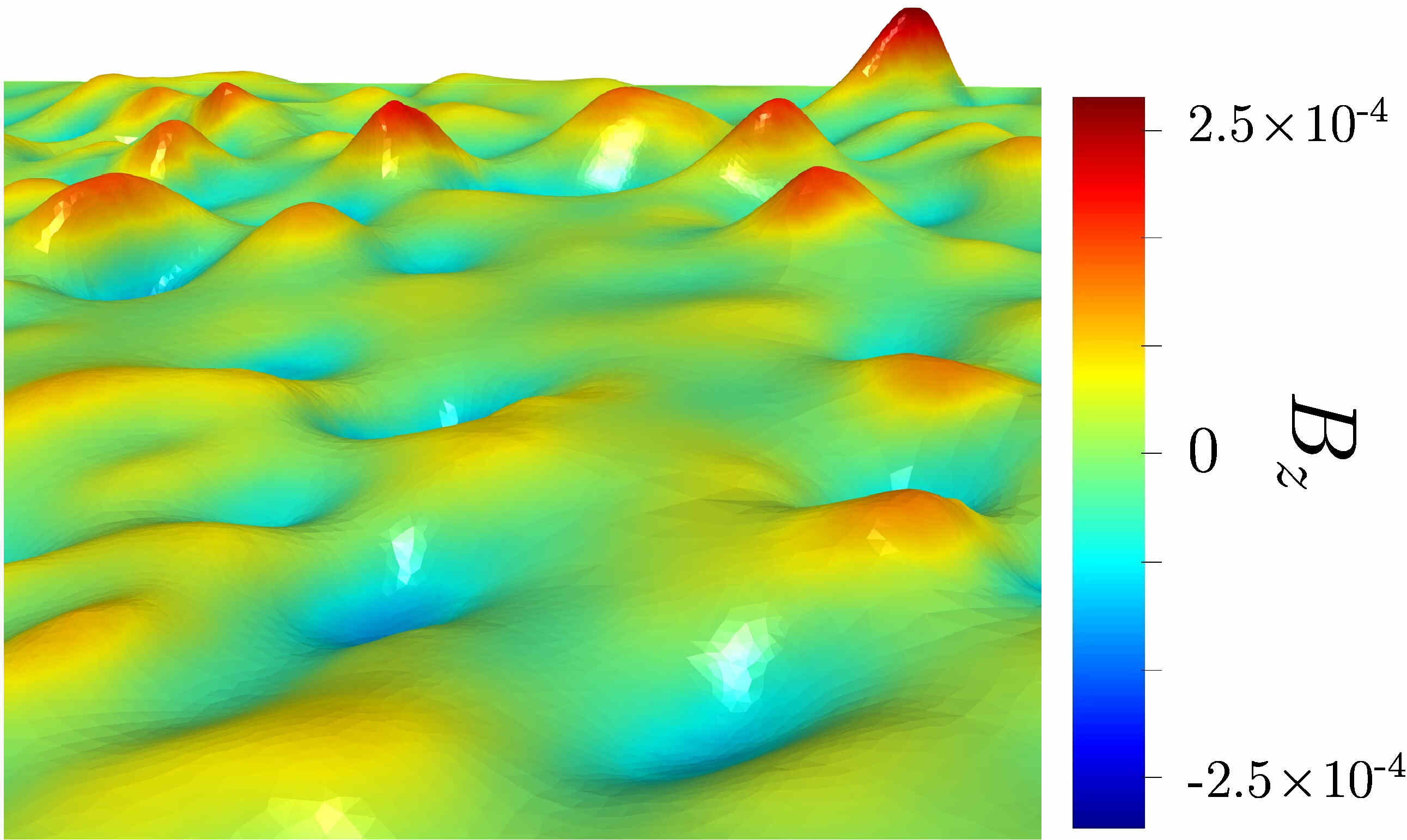}
\hss}
\caption{ 
Spontaneous magnetic field $\B$ \Eqref{Eq:M:Magnetic:1} in the quartic phase, 
generated by inhomogeneities. The inhomogeneities are modeled by random spatial 
modulation of the parameters $\alpha^\m_0$ and $\alpha^\m_z$, reflecting the naturally 
present weak gradients in doping level. 
The surface elevation, together with the coloring, represents the magnitude of the $B_z$.
The coupling here is $e=0.6$, and the other parameters are given in Supplemental 
Material \cite{Supplementary-arxiv}.
}
\label{Fig:Spontaneous}
\end{figure}


The model describing the resistive fermion quadrupling state is alike to the 
Faddeev-Skyrme model \cite{Faddeev.Niemi:97}. This suggests that it could host 
nontrivial topological excitation such as skyrmions and hopfions.
To investigate the properties of the topological defects of the effective model, 
the physical degrees of freedom $\m$ are discretized within a finite-element formulation 
\cite{Hecht:12}, and the free energy \Eqref{Eq:M:FreeEnergy:2} is minimized using a 
nonlinear conjugate gradient algorithm. For details of the numerical procedure, 
see \cite{Supplementary-arxiv}.

\begin{figure*}[!htb]
\hbox to \linewidth{ \hss
\includegraphics[width=0.9\linewidth]{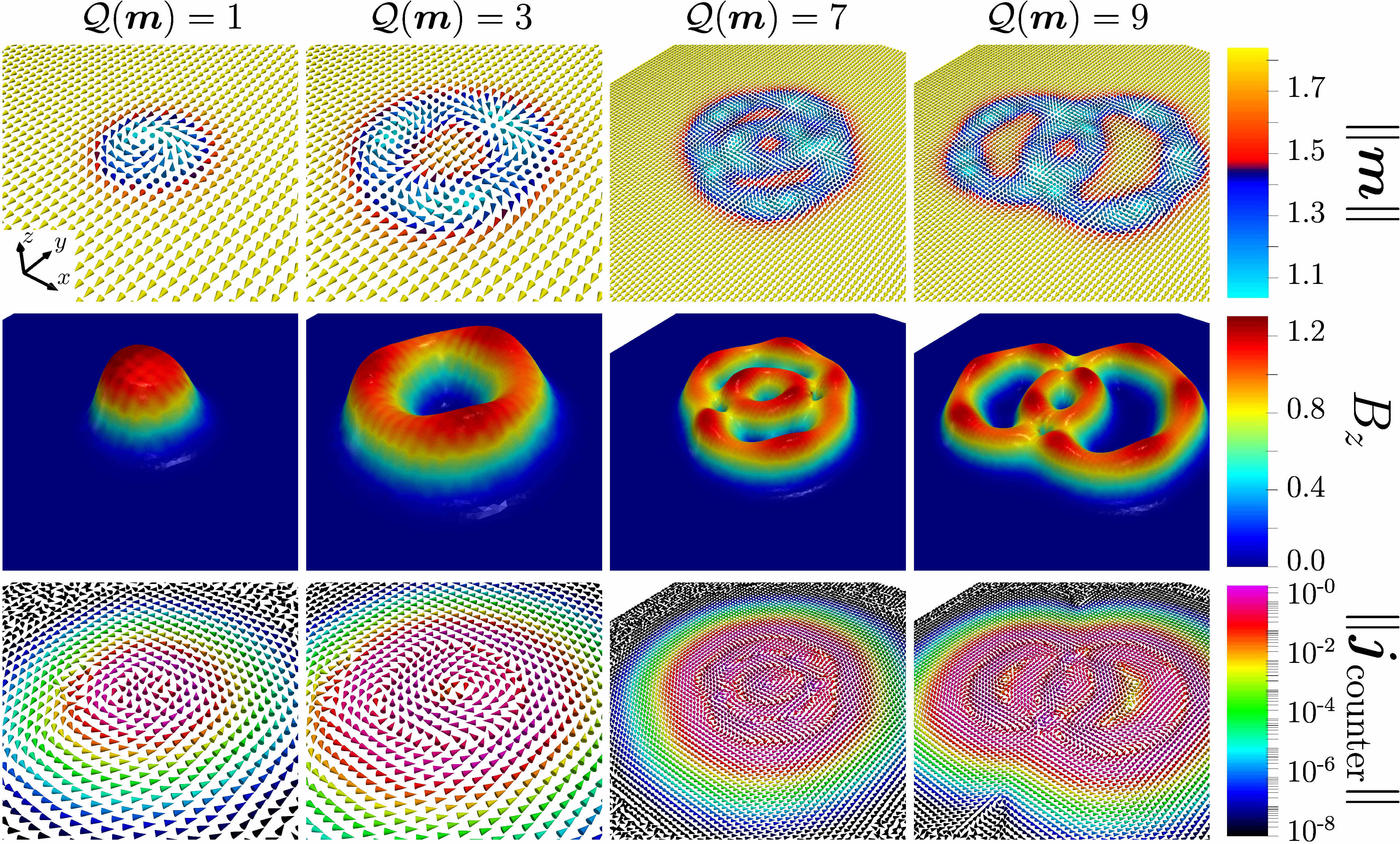}
\hss}
\caption{ 
Skyrmion solutions in a time-reversal symmetry broken state, for increasing values 
of the topological charge $\Q(\m)$. The panels on the top row display the texture of 
the four-fermion order parameter $\m$. The panels in the middle row show the associated 
magnetic field $\B$ \Eqref{Eq:M:Magnetic:1}, and the bottom row shows the corresponding 
charge transferring counter-currents $\j_\text{counter}$ according to Amp\`ere's law.
The parameters are the same as in \Figref{Fig:Spontaneous}, while the coupling $e=0.25$.
}
\label{Fig:Skyrmions}
\end{figure*}

The experiments \cite{Grinenko.Weston.ea:21} reported spontaneous magnetic fields in 
the quartic metal state. In the $\sis$ superconducting state, spontaneous magnetic 
fields can arise due to inhomogeneities such as thermal gradients 
\cite{Silaev.Garaud.ea:15,Grinenko.Weston.ea:21}, a hotspot created by a laser pulse 
\cite{Garaud.Silaev.ea:16}, the effect of impurities \cite{Maiti.Sigrist.ea:15,
Lin.Maiti.ea:16}, and other inhomogeneous arrays \cite{Garaud.Corticelli.ea:18a,
Vadimov.Silaev:18}.
The material has slight inhomogeneity in doping level, which results in relatively small 
local modulation of the superconducting critical temperature \cite{Iguchi:22}. Since for 
this topic the relative values of the gaps and phases strongly depend on doping, this 
can be modeled by spatial modulation of the prefactors of the quadratic terms of the 
Ginzburg-Landau theory. Implementing smoothly spatially varying amplitudes of the 
individual components, at the level of the effective model, can thus be modeled by 
small spatial variations of the coupling constants $\alpha^\m_0$ and $\alpha^\m_z$ 
(see Supplemental Material for details \cite{Supplementary-arxiv}). As shown in 
\Figref{Fig:Spontaneous}, such inhomogeneities in the effective model for the fermion 
quadrupling state, which breaks the time-reversal symmetry, result in spontaneous magnetic 
fields. It is qualitatively in accordance with the experiment \cite{Grinenko.Weston.ea:21}.

First note that because the time-reversal  symmetry ($\groupZ{2}$) is broken, the model 
has domain-wall excitations. These are similar, in a way, to the domain walls found in 
a three-component model \cite{Garaud.Babaev:14,Grinenko.Weston.ea:21}. They are thus 
discussed in the Supplemental Material \cite{Supplementary-arxiv}.
However the quantization of $\Q(\m)$ suggests that the model has more nontrivial 
topological excitations with quantized magnetic flux according to $\int B_z=\Phi_0\Q$, 
where $\Phi_0=-2\pi/e$ is the flux quantum. If a model breaks the $\groupZ{2}$ symmetry 
and has only gradient terms which are second order in derivatives, according to the 
Hobart-Derrick theorem \cite{Hobart:63,Derrick:64}, skyrmions cannot exist. In our case, 
the presence of the Skyrme term, in the effective model \Eqref{Eq:M:FreeEnergy:2}, 
allows for nontrivial configurations that evade the Hobart-Derrick theorem. 
Indeed, in two dimensions, the Skyrme term in the effective model scales as $1/R^2$ 
(where $R$ is a texture size), and therefore stable skyrmions may exist due to 
the competition between the Skyrme and potential terms.

We performed numerical simulation by minimizing the energy \Eqref{Eq:M:FreeEnergy:2} 
from various initial states. When the initial guess has a nontrivial topological charge, 
the minimization procedure leads, after convergence of the algorithm, to stable skyrmion 
configurations. Figure~\ref{Fig:Skyrmions} shows these skyrmions solutions for increasing 
values of the topological charge $\Q(\m)$, which is integer with an accuracy around 
$10^{-4}$. 
As shown on the middle row of \Figref{Fig:Skyrmions}, the skyrmions carry a nonzero 
magnetic field. Moreover, since the topological charge \Eqref{Eq:M:Charge} 
is quantized, the skyrmions carry integer quanta of magnetic flux.
The circulating current pattern that induces this magnetic field is illustrated in 
the bottom row. This current, defined according to Amp\`ere's law for the magnetic 
field \Eqref{Eq:M:Magnetic:1} corresponds to the charge-carrying counterflow between 
the different components.

Furthermore we find that the interskyrmion forces are attractive. Hence, single quanta 
skyrmions attract each other to form skyrmions with higher topological charge. Thus in 
general one would not expect the formation of regular skyrmion lattices but rather 
skyrmion lumps formed by the competition between the attractive forces and pinning 
landscape.
Interestingly, in a single quantum skyrmion, the time-reversed state is realized 
at a zero measure area inside the skyrmion. On the other hand, skyrmions carrying more 
than one quantum feature inner regions of the time-reversed state. The enclosed area 
of the time-reversed state increases with the topological charge. This suggests that 
if the $\groupZ{2}$ symmetry associated with the relative phase locking  is strongly 
broken, the formation of skyrmions is strongly inhibited. Note that unlike in 
\Figref{Fig:Spontaneous}, the parameters for the skyrmions displayed in \Figref{Fig:Skyrmions} 
are homogeneous, as we focus here on the detailed structure of the skyrmions. Inhomogeneities 
can however deform the skyrmions, although we find that they do not destroy skyrmions 
(see Supplemental Material \cite{Supplementary-arxiv}).


The recent experiment reported a fermion quadrupling phase in Ba$_{1-x}$K$_x$Fe$_2$As$_2$ 
\cite{Grinenko.Weston.ea:21}. In this resistive phase, there is no condensate of Cooper 
pairs, but a four-fermion condensate which breaks the $\groupZ{2}$ time-reversal symmetry.

We derived an effective model of that resistive state, starting from a microscopic 
three-band model with dominant interband interaction for Ba$_{1-x}$K$_x$Fe$_2$As$_2$ 
and by implementing a mean-field approximation for the fields that are fourth order 
in fermions. 
The effective field theory has a structure similar to the Faddeev-Skyrme model, but for 
a soft modulus vector field that represents the fermion quadrupling order parameter.
If spatial inhomogeneities are present the model accounts for spontaneous magnetic fields, 
consistently with the experimental observations \cite{Grinenko.Weston.ea:21}. 
We report that despite the lack of Meissner effect and the lack of conserved $\groupU{1}$ 
topological charge, the model has stable topological excitations in the form of skyrmions 
with conserved topological charge given by \Eqref{Eq:M:Charge}. 

We would like to remind the reader that, similarly to skyrmions that appear in other 
contexts, such as magnetism, their existence also depends on factors that are beyond the 
effective long-wavelength field-theoretic model. Namely, in contrast to vortices, the 
skyrmionic topological charge is obtained through a surface integral.
Consequently, if the terms that break the $\groupO{3}$ symmetry are very strong, 
the localization of the skyrmionic topological charge can shrink down to scales where the 
effective theory is ill defined, thereby destroying the topological protection. When the 
effective field theory is applicable, the potential barrier preventing the collapse of a 
skyrmion in a film can be roughly estimated as follows: the condensation energy density 
($F_c$) multiplied by the coherence volume $F_c\xi_{\groupZ{2}}^2L$, where $\xi_{\groupZ{2}}$ 
is the coherence length associated with the broken time-reversal symmetry and $L$ is the 
film thickness.

Finally, within the range of applicability of the effective theory, the skyrmions can be 
induced by taking advantage of the Kibble-Zurek mechanism \cite{Kibble:76,Zurek:85}, by 
quenching the material through the $\groupZ{2}$ phase transition where the time-reversal 
symmetry is broken. We expect that skyrmions may also form by cooling through the phase 
transition with an applied {\it local} magnetic field induced through a system of coils.

\begin{acknowledgments}
We thank Vadim Grinenko for discussions. 
The work was supported by the Swedish Research Council Grants 2016-06122, 2018-03659.
The computations were performed on resources provided by the Swedish 
National Infrastructure for Computing (SNIC) at the National Supercomputer 
Center at Link\"{o}ping, Sweden. 

\end{acknowledgments}

%

\setcounter{equation}{0}
\setcounter{figure}{0}
\setcounter{table}{0}
\setcounter{section}{0}
\makeatletter
\renewcommand{\theequation}{S\arabic{equation}}
\renewcommand{\thefigure}{S\arabic{figure}}
\renewcommand{\bibnumfmt}[1]{[S#1]}
\renewcommand{\citenumfont}[1]{S#1}
\onecolumngrid
\pagebreak
\begin{center}
\textbf{\large Supplemental Material: Skyrmions and magnetic properties of the 
			resistive electron quadrupling state}
\vskip 0.4cm
\begin{minipage}{0.9\textwidth}\parindent11.10839pt 
\indent 
In the Supplemental Material, we discuss the details of the derivation of the 
effective theory for the fermion quadrupling state. In particular, we start from 
the microscopic model of a three band superconductor with interband dominated 
pairing. This yields a two-component Ginzburg-Landau theory with inter-component 
mixed gradient terms which are then eliminated by a reparametrization of the 
superconducting degrees of freedom. Next, the theory is mapped to a model that 
couples the fermion quadrupling order parameter to the Meissner current. 
In the resistive state, the Meissner screening is absent, and the theory reduces 
to a model that depends only on the four-fermion order parameter.
We also discuss details of the numerical methods, and present additional results.
These include additional skyrmion solutions, the effect of material inhomogeneities 
on skyrmions, and domain-wall solutions.

\end{minipage}
\vskip 0.4cm
\end{center}
\twocolumngrid

\section{Microscopic derivation of the effective model} 
\label{Sec:App:Microscopic}

\begin{figure}[!tb]
\hbox to \linewidth{\hss
\includegraphics[width=.75\linewidth]{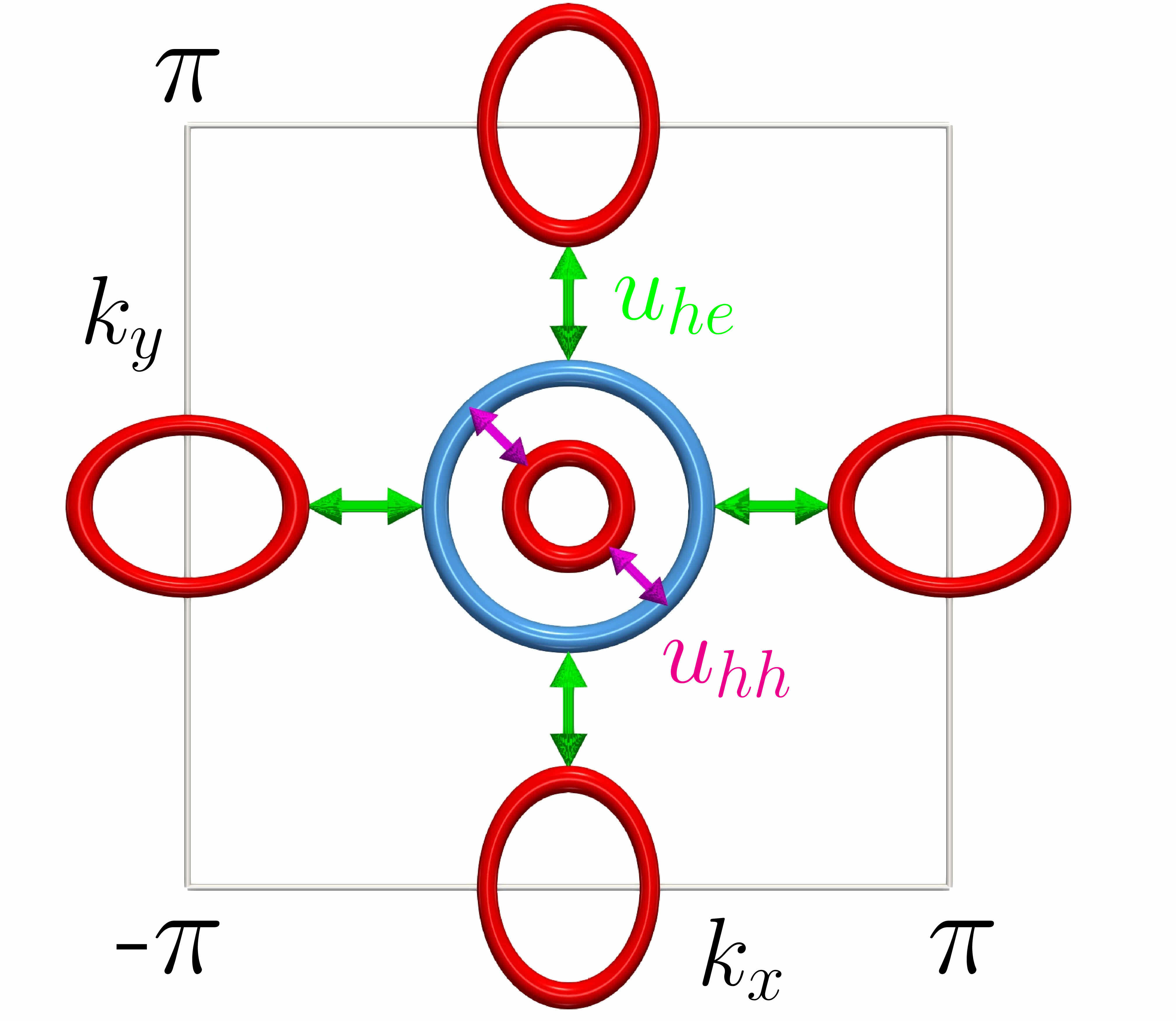}
\hss}
\caption{
Schematic view of the band structure ofr the hole-doped iron pnictide compound 
Ba$_{1-x}$K$_x$Fe$_2$As$_2$. It consists of two hole pockets at the $\Gamma$ point
shown by circles and two electron pockets at $(0; \pi)$ and $(\pi; 0)$ 
displayed by ellipses. As discussed in the text, the $\sis$ state is favoured 
by the superconducting coupling that is dominated by the interband repulsion between 
the electron and the hole Fermi surfaces $u_{he}$, as well as between the two hole 
pockets $u_{hh}$.
}
\label{Fig:App:Micro-BZ}
\end{figure}

The first part starts with the microscopic derivation of the two-component
Ginzburg-Landau theory that is relevant to describe a three-band superconductor 
with interband dominated repulsive pairing. See \cite{Supp:Garaud.Silaev.ea:17} for 
a more detailed derivation.
We are interested in values of coupling constants that can result in superconducting 
states that spontaneously break the time-reversal symmetry, aiming in particular 
to describe iron pnictides. The band structure of iron pnictides typically consists 
of two electron pockets at $(0,\pi)$ and $(\pi,0)$ and of two hole pockets at the 
$\Gamma$ point. This structure is sketched on \Figref{Fig:App:Micro-BZ}, where the 
dominating pairing channels are the interband repulsion between the two hole pockets 
at $\Gamma$, as well as between the electron and the hole bands. Note that, the order 
parameter is the same in both electron pockets, so that the crystalline $C_4$ symmetry 
is not broken and thus corresponds to an $s$-wave state.

\subsection{Generic three-component expansion}

We consider the microscopic model of a clean superconductor with three overlapping 
bands at the Fermi level. Within the quasiclassical approximation, the band parameters 
that characterize the different cylindrical sheets of the Fermi surface are the partial 
densities of states (DOS) $\nu_a$, and the Fermi velocities ${\bs v}^{(a)}_{F}$; 
here the index $a=1,2,3$ labels the different bands.
The Eilenberger equations for the quasiclassical propagators read as
\SubAlign{Eq:App:EilenbergerF}{
 &\hbar{\bs v}^{(a)}_{F}\D\phantom{^*} {f^{\phantom{+}}_a} +
 2\omega_n f_a^{\phantom{+}} - 2 \Delta_a g_a=0, \\ 
 &\hbar{\bs v}^{(a)}_{F}\D^* f^+_a -
 2\omega_n f^+_a + 2\Delta^*_a g_a=0 \,,
}
where $\omega_n = (2n+1)\pi T $, with $n\in\groupZ{}$, are the fermionic Matsubara 
frequencies and $T$ is the temperature. The gauge derivative is $\D\equiv\Grad +ie \A$, 
where $\A$ is the vector potential, and the gauge coupling is related to the flux 
quantum $\Phi_0$ by $e=-2\pi/\Phi_0$.

The quasi-classical propagators $f_a$ and $g_a$ are respectively, the anomalous 
and the normal Green's functions in each band; they obey the normalization 
condition $|f_a|^2 + g_a^2 =1$. The components $\Delta_a$ of the order parameter 
are determined by the self-consistency equations
\Equation{Eq:App:SelfConsistentGap}{
 \Delta_a ({\bs p},{\bs r})= 2\pi T \sum_{n,{\bs p^\prime},b} 
 \lambda_{ab}({\bs p},{\bs p^\prime}) f_b ({\bs p},{\bs r}, \omega_n) 
\,.
}
Here, the parameters ${\bs p}$ run over the Fermi surfaces, and $\lambda_{ab}$ are 
the components of the coupling potential matrix. For simplicity the pairing states 
are assumed to be isotropic on each of the Fermi surfaces, so that 
$\lambda_{ab}({\bs p},{\bs p^\prime} )= const$, see details in \cite{Supp:Garaud.Silaev.ea:17}. 
Finally, the self-consistent electric current is 
\Equation{Eq:App:SelfConsistentCurrent}{
 \j ({\bs r})=  2\pi e T  \sum_{n, {\bs p},a}
 \nu_j{\bs v}_F^{(a)} \Im\; g_a ({\bs p},{\bs r}, \omega_n)
}
where $\nu_a$ is the partial density of state, and 
$g_a=\mathrm{sign} (\omega_n)\sqrt{1- f_af^+_a}$.

The Ginzburg-Landau functional, is obtained by expressing the solutions of the Eilenberger 
equations \Eqref{Eq:App:EilenbergerF} as an expansion by powers of the gap functions 
amplitudes $\Delta_{a}$ and of their gradients: 
\Align{Eq:App:GLexpansion}{
  f_a&({\bs p},{\bs r}, \omega_n) = 
 \frac{\Delta_a}{\omega_n}-\frac{|\Delta_a|^2\Delta_a}{2\omega_n^3}
\\ \nonumber &-\frac{ \hbar( {\bs v}^{(a)}_{F} \D) \Delta_a }{{2\omega_n^2}}
 +\frac{\hbar^2( {\bs v}^{(a)}_{F} \D) 
 ( {\bs v}^{(a)}_{F} \D) \Delta_a}{4\omega_n^3} \,.
}
The summation over the Matsubara frequencies gives
\Equation{}{
 2\pi T \sum_{n=0}^{N_d} \omega_n^{-1} = G_0 + \tau \,,
 ~~\text{with}~~\tau = (1- T/T_c)\,.
}
Note that, $ f^+_a ({\bs p},{\bs r}, \omega_n)  = f^*_a(-{\bs p},{\bs r}, \omega_n)$.

The Ginzburg-Landau equations a determined by the substituting the expansion 
\Eqref{Eq:App:GLexpansion} into the self-consistency equation
\Eqref{Eq:App:SelfConsistentGap}. After normalizing the gaps functions by 
$T_c/\sqrt{\rho}$ (where $\rho =\sum_n \pi T_c^3\omega_n^{-3} \approx 0.1$), 
the Ginzburg-Landau equations read as
\Equation{Eq:App:GL3Band}{
  \big[(G_0+\tau-\hat\Lambda^{-1}){\bs\Delta} \big]_a
  = -K^{(a)}_{ij}D_iD_j\Delta_a + |\Delta_a|^2\Delta_a  \,,
}
where ${\bs \Delta} = (\Delta_1,\Delta_2,\Delta_3)^T$, and the anisotropy tensor is 
$K^{(a)}_{ij} =\hbar^2\rho \left\langle v^{(a)}_{Fi}v^{(a)}_{Fj} \right\rangle /2T_c^2$. 
The indices $i,j$ stand for the $x,y$ coordinates, and the average is taken over the 
$a$-th Fermi surface.
The current reads as
 \Equation{Eq:App:CurrentGL}{
 \J ({\bs r})= \frac{4e}{\hbar}\frac{T_c^2}{\rho}
 \sum_{a=1}^3 \nu_a \Im\; \Delta^*_a \hat K^{(a)} \D \Delta_a \,.
}

The critical temperature is given by the smallest positive eigenvalue of the inverse 
coupling matrix $\hat\Lambda^{-1}$. Namely, if $\lambda_{n}^{-1}$ denote the 
positive eigenvalues of the inverse coupling matrix $\hat\Lambda^{-1}$, the critical 
temperature is determined by the equation $G_0 = \min_n (\lambda^{-1}_n)$. Provided that 
all the eigenvalues are positive, the number of components of the effective field theory 
coincide with the number of bands. In this case, the Ginzburg-Landau equations for the 
three-component system read as
\Equation{Eq:App:GL3component}{
   -K^{(a)}_{ij}D_iD_j \Delta_a + \alpha_{aa}\Delta_a 
   + \alpha_{ab}\Delta_b + \beta_a|\Delta_a|^2\Delta_a =0,
}
where  
\SubAlign{Eq:App:GL3componentCoef}{ 
\alpha_{aa} = (\hat\Lambda^{-1}_{ab} - G_0-\tau)\delta_{ab}, \\
 \alpha_{ab} = (1-\delta_{ab}) \hat\Lambda^{-1}_{ab} 
 ~~~\text{and}~~~ \beta_a = 1 \,.
} 

While the precise microscopic physics behind the superconductivity in 
Ba$_{1-x}$K$_x$Fe$_2$As$_2$ is still unknown, we focus on the scenario of a three-band 
model with interband dominated repulsive pairing. In this case, the eigenvalues of the 
inverse coupling matrix are not all positive. This implies, as detailed below, that the 
three-band theory is described by a two-component order parameter.

\subsection{Two-component Ginzburg-Landau theory for the \texorpdfstring{$\sis$}{s+is} 
superconducting state}

Our principal interest here, is the time-reversal symmetry breaking $\sis$ state 
in a three-band superconductor. We consider an interband dominated repulsive pairing, 
suggested to be relevant for iron-based superconductors \cite{Supp:Maiti.Chubukov:13}. 
The corresponding coupling matrix $\hat\Lambda$ is parametrized as
\Equation{Eq:App:Model3BandB1}{
\hat\Lambda = - \left(%
\begin{array}{ccc}
  0        & u_{hh}    & u_{eh}   \\
  u_{hh}     & 0       & u_{eh}   \\
  u_{eh}  & u_{eh} & 0         \\
\end{array}  \right) \,.
}
Thus the fields $\Delta_{1,2}$ correspond to the gap functions at the hole Fermi surfaces 
while $\Delta_3$ is the gap at the electron pockets sketched in \Figref{Fig:App:Micro-BZ}. 
The coefficients $u_{hh}$ and $u_{eh}$ are respectively the hole-hole and electron-hole 
interactions. 
The linear equation that determines the critical temperature $G_0=\min(G_1,G_2)$ is 
obtained by neglecting the r.h.s. of \Eqref{Eq:App:GL3Band}. Here $G_1$ and $G_2$ 
are the only two positive eigenvalues of the inverse coupling matrix
\Equation{Eq:App:Model3BandB2}{
 \hat\Lambda^{-1} =  \frac{1}{2u_{eh}^2u_{hh}}\left(
 \begin{array}{ccc}
  u_{eh}^2     & -u_{eh}^2    & -u_{eh}u_{hh}  \\
  -u_{eh}^2    & u_{eh}^2     & -u_{eh}u_{hh}  \\
  -u_{eh}u_{hh}  & -u_{eh}u_{hh}  & u_{hh}^2        \\
\end{array}  \right) \,.
}
They explicitly reads as $G_1 = 1/u_{hh}$ and 
$G_2 =\left(u_{hh}+\sqrt{u_{hh}^2+8u_{eh}^2} \right) /4u_{eh}^2$.
The associated eigenvectors are ${\bs\Delta}_1=(-1,1,0)^T$ and ${\bs\Delta}_2=(x,x,1)^T$, 
where $x=(u_{hh}-\sqrt{u_{hh}^2+8u_{eh}^2})/4u_{eh}$. Since the only fields that can 
nucleate are those associated with positive eigenvalues, the Ginzburg-Landau theory 
\Eqref{Eq:App:GL3Band} has to be reduced to a two-component one. This reduction is obtained 
by expressing the general order parameter as the linear combination
\Align{Eq:App:OProtation}{
 {\bs \Delta}&=  \eta_1 {\bs \Delta}_1 + \eta_2 {\bs \Delta}_2\,, \nonumber \\
 \text{and}~~~(\Delta_1,\Delta_2,\Delta_3)&=(x\eta_2-\eta_1,x\eta_2+\eta_1,\eta_2)\,.
}
Here $\eta_1$ and $\eta_2$ are the order parameter of the $s_\pm$ pairing channels 
respectively between the two concentric hole surfaces and between the hole and electron 
surfaces.

The substitution of the linear combination \Eqref{Eq:App:OProtation} into the 
Ginzburg-Landau equations \Eqref{Eq:App:GL3Band}, after projection onto the 
eigenvectors ${\bs \Delta}_{1,2}$, yields the system of two Ginzburg-Landau 
equations \cite{Supp:Garaud.Silaev.ea:17}:
\SubAlign{Eq:App:GL3BandReduced}{
 a_{11}\eta_1 + b_{1j}|\eta_j|^2\eta_1 + c_{12}\eta_1^*\eta_2^2 & = 
\frac{k_{1j}}{2} \D\D  \eta_j  \,,
\\
 a_{22}\eta_2 + b_{2j}|\eta_j|^2\eta_2 + c_{12}\eta_2^*\eta_1^2 &= 
\frac{k_{2j}}{2} \D\D  \eta_j  \,.
}
The parameters on the left hand side of the Ginzburg-Landau equations 
\Eqref{Eq:App:GL3BandReduced} are expressed, in terms of the coefficients 
of the coupling matrix \Eqref{Eq:App:Model3BandB1} as 
\Align{Eq:App:Parameters}{
a_{jj} &= -|\bs \Delta_j|^2(G_0-G_{j}+\tau )\,,~~a_{12}=0 \\
b_{11} &= 2    \,,~~b_{22}=(2x^4+1) \,,~~ b_{12} = 4x^2 \,,~c_{12}=2x^2 
\,, \nonumber
}
where $|\bs \Delta_1|^2 =2 $ and $|\bs \Delta_2|^2=2x^2+1$. The $s+is$ state is 
symmetric under the $C_4$ transformations, thus the coefficients satisfy 
$K^{(j)}_{xx} = K^{(j)}_{yy} = K^{(j)}$. As a results, the coefficients of the 
gradient terms in \Eqref{Eq:App:GL3BandReduced} read as 
\SubAlign{Eq:App:GradientCoeffitients}{
   k_{11} & = 2\xi_0^{-2}\big[ K^{(1)} +  K^{(2)} \big]				\\
   k_{22} & = 2\xi_0^{-2}\big[(K^{(1)} + K^{(2)})x^2 + K^{(3)} \big]	\\
   k_{12} & = 2\xi_0^{-2} x\big[K^{(2)} - K^{(1)} \big] \,.
}

The total superconducting current \Eqref{Eq:App:CurrentGL}, is the superposition of 
the partial currents $\J^{(a)}$ of the different components of the order parameters, 
as $\J=\sum_a\J^{(a)}$; and the partial currents read as 
\Equation{Eq:App:Currents}{
\J^{(a)} = e\Im\big(\eta_a^*\sum_b\left[k_{ab}\D\eta_b\right]\big) \,.
}
The two-component free energy functional that corresponds to the 
Ginzburg-Landau equations \Eqref{Eq:App:GL3BandReduced}, and whose variations 
with respect to $\A$ give the supercurrent \Eqref{Eq:App:Currents}, reads as 
(in dimensionless units): 
\SubAlign{Eq:App:FreeEnergy}{
\F =\frac{\B^2}{2}+&\frac{1}{2}\sum_{a,b=1}^2
 	k_{ab}(\D\eta_a)^*\D\eta_b
 	+V({\bs\eta}) \,,	\\
\text{where}~V({\bs\eta}) &= \sum_{a,b=1}^2a_{ab}\eta_a^*\eta_b
	+\frac{b_{ab}}{2}|\eta_a|^2|\eta_b|^2
\\
&~~~+\frac{c_{12}}{2}\big(\eta_1^{*2}\eta_2^2+c.c.\big)
		\,.
}
Here, the complex fields $\eta_{1,2}$ are the components of the superconducting order 
parameter. They are electromagnetically coupled by the vector potential $\A$ of the 
magnetic field $\B=\Curl\A$, through the gauge derivative $\D\equiv\Grad+ie\A$. 
There, the coupling constant $e$ is used to parametrize the London penetration length.
Note that for the energy to be positive definite, the coefficients of the kinetic terms 
should satisfy the relation $\det\hk=k_{11}k_{22}-k_{12}^2>0$. Also, for the free 
energy functional to be bounded from below, the coefficients of the terms that are fourth 
order in the condensates should satisfy the condition $b_{11}b_{22}-(b_{12}+c_{12})^2>0$. 
Finally, the condition for having a nonzero ground-state density is 
$\det\hat{a}=a_{11}a_{22}-a_{12}^2<0$. These conditions are of course satisfied by 
the microscopically calculated value \Eqref{Eq:App:Parameters} and 
\Eqref{Eq:App:GradientCoeffitients}.

\subsection{Elimination of the mixed-gradients by diagonalization}
\label{Sec:App:Mixed-gradients}

Within the current basis for the superconducting degrees of freedom, it is quite 
complicated to deal with the kinetic terms. It is thus worth rewriting the model using 
a linear combination of the components of the order parameter, that diagonalize the 
kinetic term:
\Equation{Eq:App:FreeEnergy:Kinetic:1}{
\F_k=\frac{1}{2}\sum_{a,b=1}^2 k_{ab}(\D\eta_a)^*\D\eta_b:=
\frac{1}{2}(\D{\bs\eta})^\dagger\hk\D{\bs\eta}	\,.
}
Here ${\bs\eta}^\dagger=(\eta_1^*,\eta_2^*)$, and $\hk$ is the matrix whose 
elements are $k_{ab}$. The positive definiteness of the free energy implies that 
$\det\hk>0$. So, $\hk$ is a positive definite square matrix whose square 
root is
\Equation{Eq:App:Kinetic:SquareRoot}{
\cR = \frac{\hk+\Identity\sqrt{\det\hk}}{\sqrt{\tr\hk+2\sqrt{\det\hk}} }
\,,~\text{where}~\hk= \cR^\dagger\cR \,.
}
This determines a natural linear combination of the superconducting degrees of freedom, 
where the kinetic term \Eqref{Eq:App:FreeEnergy:Kinetic:1} is diagonal: 
\Equation{Eq:App:FreeEnergy:Kinetic:2}{
\F_k=\frac{1}{2}(\D\Psi)^\dagger\D\Psi	
\,,~\text{where}~\Psi=\cR{\bs\eta} \,,
}
and $\Psi^\dagger=(\psi_1^*,\psi_2^*)$.
The original superconducting degrees of freedom ${\bs\eta}$ are restored via the 
reverse transformation ${\bs\eta}=\cR^{-1} \Psi$, where
\Equation{Eq:App:Kinetic:SquareRoot:Inverse}{
\cR^{-1}=\frac{\sqrt{\tr\hk+2\sqrt{\det\hk}}}{\det(\hk+\Identity\sqrt{\det\hk})}
\big[ (\tr\hk+\sqrt{\det\hk})\Identity-\hk  \big]	\,.
}

Using the relations \Eqref{Eq:App:Kinetic:SquareRoot} and 
\Eqref{Eq:App:Kinetic:SquareRoot:Inverse} to parametrize the superconducting degrees 
of freedom with $\Psi$ instead of ${\bs\eta}$, greatly simplifies the kinetic term 
\Eqref{Eq:App:FreeEnergy:Kinetic:2}. Thus, the potential term in the free energy  
\Eqref{Eq:App:FreeEnergy} has to be rewritten in terms $\Psi$. In all generality, 
the potential energy reads as 
\Equation{Eq:App:FreeEnergy:Potential:1}{
\F_p :=V({\bs\eta})= a_{ij} \eta_i^*\eta_j^{}
+ \frac{b_{ijkl}}{2} \eta_i^*\eta_j^*\eta_k^{}\eta_l^{}	\,,
}
with the summation over the repeated indices. Note that for the energy to be a real 
quantity, the tensor coefficients $a_{ij}$, and $b_{ijkl}$ should obey some symmetry 
relations: 
\SubAlign{Eq:App:FreeEnergy:Potential:symmetry}{
 a_{ij}    &=  a_{ji} \,\\
 b_{ijkl}  &=  b_{jikl} = b_{ijlk} = b_{klij}
 \,.
}
Similarly, in terms of $\Psi$ the potential energy reads as
\Equation{Eq:App:FreeEnergy:Potential:2}{
\F_p :=V({\Psi})= \alpha_{ij} \psi_i^*\psi_j^{}
+ \frac{\beta_{ijkl}}{2} \psi_i^*\psi_j^*\psi_k^{}\psi_l^{}	\,,
}
where the tensor coefficients $\alpha_{ij}$ and $\beta_{ijkl}$ obey the same symmetry 
relations \Eqref{Eq:App:FreeEnergy:Potential:symmetry} as $a_{ij}$, and $b_{ijkl}$.
They are obtained via the transformation ${\bs\eta}=\cR^{-1} \Psi$, and the relations 
are 
\SubAlign{Eq:App:FreeEnergy:Potential:transfromation}{
 \alpha_{ij}    &=  a_{ab}\cR^{-1}_{ai}\cR^{-1}_{bj} \,\\
 \beta_{ijkl}  &=  b_{abcd}\cR^{-1}_{ai}\cR^{-1}_{bj}\cR^{-1}_{ck}\cR^{-1}_{dl}
 \,.
}
\begin{widetext}
A simple, yet lengthy algebraic manipulations thus yield the free energy in terms of the 
superconducting degrees of freedom $\Psi$
\SubAlign{Eq:App:FreeEnergy:diagonal}{
\F &=\frac{\B^2}{2}+\frac{1}{2}(\D\Psi)^\dagger\D\Psi
 	+V(\Psi) \,,~~\text{where}~\B=\Curl\A \,,~~\D\equiv\Grad+ie\A\,,	\\
\text{and}~V(\Psi) = \sum_{i,j=1}^2\alpha_{ij}\psi_i^*\psi_j
	&+\frac{\beta_{ij}}{2}|\psi_i|^2|\psi_j|^2  	
	+\big(\gamma_{11}|\psi_1|^2+\gamma_{22}|\psi_2|^2\big)\big(\psi_1^{*}\psi_2+c.c.\big) 
+\frac{\gamma_{12}}{2}\big(\psi_1^{*2}\psi_2^2+c.c.\big)
		\,,
}
and $\beta_{ij}:=\beta_{ijkl}\delta_{ki}\delta_{jl}$,  
$\gamma_{ik}=\beta_{ijkl}\delta_{j1}\delta_{l2}$. Moreover, all the coefficients 
are symmetric, for example $\beta_{21}=\beta_{12}$.
Using the relations \Eqref{Eq:App:FreeEnergy:Potential:transfromation}, and collecting 
the various terms yields the relation for the coefficients of the bilinear terms
\SubAlign{Eq:App:Coefficients:bilin}{
\alpha_{11} &= a_{11} (\cR^{-1}_{11})^2 + a_{22} (\cR^{-1}_{21})^2  
		    + 2a_{12} \cR^{-1}_{11} \cR^{-1}_{21} \\
\alpha_{22} &= a_{11} (\cR^{-1}_{12})^2 + a_{22} (\cR^{-1}_{22})^2  
			+ 2a_{12}\cR^{-1}_{12} \cR^{-1}_{22}           \\
\alpha_{12} &= a_{11} \cR^{-1}_{11}\cR^{-1}_{12} 
			 + a_{22}\cR^{-1}_{21}\cR^{-1}_{22} 
			 + a_{12}(\cR^{-1}_{11} \cR^{-1}_{22} + \cR^{-1}_{12} \cR^{-1}_{21})
			 \,.
}
Similarly, the coefficients for the fourth order terms are
\SubAlign{Eq:App:Coefficients:quart:1}{
\beta_{11} &= 
  \frac{b_{11}}{2} (\cR^{-1}_{11})^4 + \frac{b_{22}}{2} (\cR^{-1}_{21})^4 
  + (b_{12} + c_{12})(\cR^{-1}_{11})^2 (\cR^{-1}_{21})^2 
 + 2 \cR^{-1}_{11} \cR^{-1}_{21} ( c_{11}(\cR^{-1}_{11})^2 +c_{22} (\cR^{-1}_{21})^2) 
   \\
\beta_{22} &= 
  \frac{b_{11}}{2} (\cR^{-1}_{12})^4 + \frac{b_{22}}{2} (\cR^{-1}_{22})^4  
  + (b_{12} + c_{12})(\cR^{-1}_{12})^2 (\cR^{-1}_{22})^2 
  + 2\cR^{-1}_{12}\cR^{-1}_{22} ( c_{11} (\cR^{-1}_{12})^2 +c_{22}(\cR^{-1}_{22})^2)
   \\
\beta_{12} &= 
	2b_{11} (\cR^{-1}_{11})^2 (\cR^{-1}_{12})^2 
	+ 2b_{22} (\cR^{-1}_{21})^2 (\cR^{-1}_{22})^2  
   + b_{12} (\cR^{-1}_{11} \cR^{-1}_{22} + \cR^{-1}_{12} \cR^{-1}_{21})^2 
  + 4c_{12}\cR^{-1}_{11} \cR^{-1}_{12} \cR^{-1}_{21} \cR^{-1}_{22}
\nonumber \\
  &+ 4(c_{11} \cR^{-1}_{11} \cR^{-1}_{12} +c_{22} \cR^{-1}_{21} \cR^{-1}_{22} )
  (\cR^{-1}_{11} \cR^{-1}_{22} + \cR^{-1}_{12} \cR^{-1}_{21}) 
\,,   
}
and 
\SubAlign{Eq:App:Coefficients:quart:2}{
\gamma_{11} &= 
  b_{11}(\cR^{-1}_{11})^3 \cR^{-1}_{21} + b_{22}(\cR^{-1}_{12})^3 \cR^{-1}_{22}  
  + (b_{12} + c_{12}) \cR^{-1}_{11} \cR^{-1}_{21} 
   (\cR^{-1}_{11} \cR^{-1}_{22} + \cR^{-1}_{12} \cR^{-1}_{21}) 
\nonumber \\
	&+ c_{11}(\cR^{-1}_{11})^2 (\cR^{-1}_{11} \cR^{-1}_{22} + 3 \cR^{-1}_{12} \cR^{-1}_{21}) 
	+c_{22}(\cR^{-1}_{12})^2(3\cR^{-1}_{11} \cR^{-1}_{22}+\cR^{-1}_{12} \cR^{-1}_{21}) 
   \\
\gamma_{22} &= 
  b_{11}(\cR^{-1}_{21})^3 \cR^{-1}_{11} + b_{22}(\cR^{-1}_{22})^3 \cR^{-1}_{12}  
  + (b_{12}+c_{12})\cR^{-1}_{22}\cR^{-1}_{12}
  	(\cR^{-1}_{11}\cR^{-1}_{22} + \cR^{-1}_{12} \cR^{-1}_{21}) 
\nonumber \\
	&+ c_{11}(\cR^{-1}_{12})^2(3\cR^{-1}_{11}\cR^{-1}_{22}+\cR^{-1}_{12}\cR^{-1}_{21}) 
	+ c_{22}(\cR^{-1}_{22})^2(\cR^{-1}_{11}\cR^{-1}_{22}+3\cR^{-1}_{12}\cR^{-1}_{21}) 
   \\
\gamma_{12} &= 
	\frac{b_{11}}{2} (\cR^{-1}_{11})^2 (\cR^{-1}_{12})^2 
  + \frac{b_{22}}{2} (\cR^{-1}_{21})^2 (\cR^{-1}_{22})^2  
  + b_{12} \cR^{-1}_{11} \cR^{-1}_{12} \cR^{-1}_{21} \cR^{-1}_{22} 
  + \frac{c_{12}}{2}((\cR^{-1}_{11} \cR^{-1}_{22})^2 + (\cR^{-1}_{12} \cR^{-1}_{21})^2) 
\nonumber \\
	&
	+ (c_{11}\cR^{-1}_{11}\cR^{-1}_{12}+ c_{22}\cR^{-1}_{21}\cR^{-1}_{22})
	(\cR^{-1}_{11}\cR^{-1}_{22}+\cR^{-1}_{12} \cR^{-1}_{21}) 
\,.
}

\end{widetext}

Note that the elimination of the mixed gradient terms via the decomposition of the 
matrix $\hk$ in terms of the square root matrix $\cR$ is not unique. Indeed, there 
exist different possibilities, see for example \cite{Supp:Garaud.Silaev.ea:17,
Garaud.Corticelli.ea:18a}.

\subsection{Separation of charged and neutral modes} 
\label{Sec:App:Separation}

The total Meissner current $\J$ is defined by the variation of the free energy 
\Eqref{Eq:App:FreeEnergy:diagonal} with respect to the vector potential: 
\Equation{Eq:App:EOM:Current}{
\J:=\frac{\delta\F}{\delta\A}=e^2\Psi^\dagger\Psi\A
							+e\Im\big(\Psi^\dagger\Grad\Psi\big)\,.
}
It follows that the gauge field can be explicitly eliminated by expressing $\A$ 
in terms of the condensate $\Psi$, and the Meissner current:
\Equation{Eq:App:GaugeField}{
e\A= \frac{1}{e\varrho^2}\Big(\J-e\Im\big(\Psi^\dagger\Grad\Psi\big)\Big)
\,,~\text{where}~\varrho^2=\Psi^\dagger\Psi \,.
}
Indeed, the kinetic term can be written as
\Equation{Eq:App:Kinetic}{
|\D\Psi|^2 =\frac{\J^2}{e^2\varrho^2}+\Grad\Psi^\dagger\cdot\Grad\Psi
+\frac{\big(\Psi^\dagger\Grad\Psi-\Grad\Psi^\dagger\Psi\big)^2}{4\varrho^2}\,,
}
and that the magnetic field 
\SubAlign{Eq:App:MagneticField:1}{
B_k &=\epsilon_{kij}\left\lbrace 
\nabla_i\left(\frac{J_j}{e^2\varrho^2}\right) 
+\frac{i}{e\varrho^4}Z_{ij} \right\rbrace
\\
\text{where}~Z_{ij}&=\varrho^2\Grad_i\Psi^\dagger\Grad_j\Psi
+(\Psi^\dagger\Grad_i\Psi)(\Grad_j\Psi^\dagger\Psi) \,. \label{Eq:Zij}
}
Hence, the magnetic field features a contribution from the Meissner current $\J$, 
together with a contribution from the interband counterflow $Z_{ij}$. Note that 
since $Z^*_{ij}=Z_{ji}$, the magnetic field can be written as 
\Equation{Eq:App:MagneticField:2}{
B_k =\epsilon_{kij}\left\lbrace 
\nabla_i\left(\frac{J_j}{e^2\varrho^2}\right) 
-\frac{\Im Z_{ij}}{e\varrho^4} \right\rbrace \,.
}
It follows that the free energy \Eqref{Eq:App:FreeEnergy:diagonal} can be rewritten as 
\Align{Eq:App:FreeEnergy:1}{
&\F= \frac{1}{2}\left[ \epsilon_{kij}\left\lbrace 
\nabla_i\left(\frac{J_j}{e^2\varrho^2}\right) +\frac{i}{e\varrho^4}Z_{ij} 
 \right\rbrace\right]^2  + \frac{\J^2}{2e^2\varrho^2} \nonumber\\
&
+\Grad\Psi^\dagger\!\cdot\!\Grad\Psi
+\frac{1}{4\varrho^2}\big(\Psi^\dagger\Grad\Psi-\Grad\Psi^\dagger\Psi\big)^2
+ V(\Psi)	\,,
}
where $Z_ {ij}$ is defined in \Eqref{Eq:Zij}.

\subsection{Mapping to the effective model}
\label{Sec:App:Mapping}

Two-component Ginzburg-Landau models can often be mapped onto a version of the nonlinear 
$\groupO{3}$ $\sigma$-model \cite{Supp:Babaev.Faddeev.ea:02,Supp:Babaev:09,
Supp:Garaud.Sellin.ea:14,Supp:Garaud.Silaev.ea:17}. In those mappings the $\groupO{3}$ 
symmetry is explicitly broken by the potential terms, consistently with the symmetry of 
the superconducting state.
The mapping couples the massive $\groupU{1}$ vector ﬁeld (the current $\J$) to a compact 
$\groupO{3}$ unit vector (the pseudo-spin $\vn$) and a real scalar (the density $\varrho$). 
The pseudo-spin unit is defined by projecting the superconducting degrees of freedom
onto the spin-1/2 Pauli matrices $\sigma$. For derivation of this mapping for different 
two-component Ginzburg-Landau models, see \eg 
\cite{Supp:Garaud.Sellin.ea:14,Supp:Garaud.Silaev.ea:17}.

Here, we use an alternative mapping to a model that couples the massive $\groupU{1}$ 
vector field (the current $\J$) to the fermion quadrupling order parameter in the form 
of a 3-vector $\m$. The fermion quadrupling field is defined as the projection of the 
superconducting degrees of freedom $\Psi$ onto the spin-1/2 Pauli matrices ${\bs \sigma}$: 
\Equation{Eq:App:M}{
 \m\equiv (m_x,m_y,m_z)
 =\Psi^\dagger{\bs \sigma}\Psi\, .
}
Unlike the pseudo-spin $\vn$, which a unit vector, the norm of $\m$ is not fixed. 
Thus $\vn$ and $\m$ are related to each other according to $\m=\varrho^2\vn$, and we 
sometime refer to $\m$ as a soft modulus vector field. The projection \Eqref{Eq:App:M} 
determines the following relations 
\Align{Eq:App:M:Relation:1}{
\partial_i\m\cdot&\partial_j\m = 
	2\rho^2\big(\partial_i\Psi^\dagger\partial_j\Psi 
			  + \partial_j\Psi^\dagger\partial_i\Psi \big) \nonumber \\
	&+\big(\Psi^\dagger\partial_i\Psi - \partial_i\Psi^\dagger\Psi \big)
	  \big(\Psi^\dagger\partial_j\Psi - \partial_j\Psi^\dagger\Psi \big) \,,
}
where, for the product of Pauli matrices, we used the Fierz identity 
\Equation{Eq:App:Pauli:1}{
\sigma^\alpha_{ab}\sigma^\alpha_{cd}=2\delta_{ad}\delta_{bc}-\delta_{ab}\delta_{cd} \,,
}
where $\delta_{ab}$ is the Kronecker symbol. It follows that
\Equation{Eq:App:M:Relation:2}{
\big(\Grad\m\big)^2 = 	4\rho^2\Grad\Psi^\dagger\cdot\Grad\Psi 
	+\big(\Psi^\dagger\Grad\Psi - \Grad\Psi^\dagger\Psi \big)^2  \,.
}
The kinetic term \Eqref{Eq:App:Kinetic} can thus be written as
\Equation{Eq:App:M:Kinetic}{
|\D\Psi|^2 =\frac{\J^2}{e^2\varrho^2} + \frac{1}{4\varrho^2}\big(\Grad\m\big)^2 \,.
}
Similarly, the projection \Eqref{Eq:App:M} determines the relation 
\Align{Eq:App:M:Relation:3}{
&\m\cdot\partial_i\m\times\partial_j\m = -2i\varrho^4		
	\big(\partial_i\Psi^\dagger\partial_j\Psi 
	   - \partial_j\Psi^\dagger\partial_i\Psi \big) \nonumber \\
	&-2i\varrho^2\big[  (\Psi^\dagger\partial_i\Psi)(\partial_j\Psi^\dagger\Psi)
					  - (\partial_i\Psi^\dagger\Psi)(\Psi^\dagger\partial_j\Psi)
	 \big] \,,
}
where, for the triple product of Pauli matrices, we used the identity 
\Equation{Eq:App:Pauli:2}{
\epsilon_{\alpha\beta\gamma}\sigma^\alpha_{ab}\sigma^\beta_{cd}\sigma^\gamma_{ef}=2i\big(
\delta_{af}\delta_{bc}\delta_{de}-\delta_{ad}\delta_{cf}\delta_{be}	\big) \,.
}
There $\delta_{ab}$ is the Kronecker symbol, and $\epsilon_{\alpha\beta\gamma}$ is the 
rank-3 Levi-Civita symbol. It follows that 
\Align{Eq:App:M:Relation:4}{
&\epsilon_{ijk}\m\cdot\partial_i\m\times\partial_j\m = \nonumber \\
&-4i\varrho^2 \epsilon_{ijk}\Big\lbrace \varrho^2 \partial_i\Psi^\dagger\partial_j\Psi  
	+  (\Psi^\dagger\partial_i\Psi)(\partial_j\Psi^\dagger\Psi)
	 \Big\rbrace \,.
}
Hence, the magnetic field reads as
\Equation{Eq:App:M:Magnetic}{
\B = \Curl\!\left(\frac{\J}{e^2\varrho^2}\right) 
	-\frac{\epsilon_{\alpha\beta\gamma}}{4e\varrho^6}
	m_\alpha\Grad m_\beta\times\Grad m_\gamma	\,.
}
As a result, 
\Align{Eq:App:M:FreeEnergy:1}{
&\F= \frac{1}{2}\left[ \epsilon_{kij}\left\lbrace 
\nabla_i\left(\frac{J_j}{e^2\varrho^2}\right) -\frac{1}{4e\varrho^6}
	\m\cdot\partial_i\m\times\partial_j\m \right\rbrace\right]^2\nonumber\\
& + \frac{\J^2}{2e^2\varrho^2} + \frac{1}{8\varrho^2}\big(\Grad\m\big)^2
+ V(\m)	\,,
}
where the density is $\varrho^2\equiv\Norm{\m}=\sqrt{\m\cdot\m}$. 

The effective model of the fermion quadrupling resistive state is deduced by removing 
the superconducting degrees of freedom from \Eqref{Eq:App:M:FreeEnergy:1}, since their 
prefactors are renormalized to zero. Namely, in that resistive state the Meissner current 
vanishes ($\J=0$), while the currents associated with the gradients of the fermion 
quadrupling order parameter $\m$ do not. 
Assuming that the critical temperatures of $\groupZ{2}$ and $\groupU{1}$ transitions 
are well separated, and assuming a mean-field approximation for the fields that are 
fourth-order in fermions the free energy of the fermion quadrupling state reads as 
\Equation{Eq:App:M:FreeEnergy:2}{
\F=  \frac{\big(\m\cdot\partial_i\m\times\partial_j\m  \big)^2}{16e^2\Norm{\m}^6}
  + \frac{\big(\Grad\m\big)^2}{8\Norm{\m}}	+ V(\m)	\,.
}
Similarly, the magnetic field in the fermion quadrupling resistive state becomes 
\Equation{Eq:App:M:Magnetic:1}{
\B = -\frac{\epsilon_{\alpha\beta\gamma}
	m_\alpha\Grad m_\beta\times\Grad m_\gamma  }{4e\Norm{\m}^3}	\,.
}
Equivalently, component-wise, the magnetic field is 
\Equation{Eq:App:M:Magnetic:2}{
B_k = -\frac{\epsilon_{kij}
	\m\cdot\nabla_i \m\times\nabla_j \m  }{4e\Norm{\m}^3}	\,.
}

Finally, the potential term reads as
\Align{Eq:App:M:Potential}{
V(\m)= \sum_{a=0,x,y,z}\left(\alpha^\m_a  +
	   \sum_{b=0,x,y,z}\frac{\beta^\m_{ab}}{2} m_b \right) m_a\,,
}
where the component $m_0:=\Norm{\m}$, and the coefficients 
depend on the coefficients \Eqref{Eq:App:Coefficients:bilin},  
\Eqref{Eq:App:Coefficients:quart:1} and \Eqref{Eq:App:Coefficients:quart:2} of the 
diagonalized free-energy \Eqref{Eq:App:FreeEnergy:diagonal}. All the coefficients 
involving a $y$ index vanish: $\alpha^\m_y=\beta^\m_{ay}=\beta^\m_{ya}=0$.
The non-zero coefficients $\alpha^\m_a $ of the linear term in $m_a$ are 
\Equation{Eq:App:M:Coefficients:1}{
\alpha^\m_0=\frac{\alpha_{11}+\alpha_{22}}{2} \,,~ 
\alpha^\m_x= \alpha_{12}    \,,~ 
\alpha^\m_z=\frac{\alpha_{11}-\alpha_{22}}{2}     \,.
}
Next, the non-zero coefficients $\beta^\m_{ab}$ of the bilinear term in $m_a$ are 
\SubAlign{Eq:App:M:Coefficients:2}{
\beta^\m_{00}&=\frac{\beta_{11}+\beta_{22} +2(\beta_{12}-\gamma_{12}) }{4} \,,~ \\
\beta^\m_{0x}=\beta^\m_{x0}&= \frac{\gamma_{11}+\gamma_{22}}{2}    \,,~ \\
\beta^\m_{0z}=\beta^\m_{z0}&= \frac{\beta_{11}-\beta_{22}}{2}      \,,
}
and
\SubAlign{Eq:App:M:Coefficients:3}{
\beta^\m_{xx}&=\gamma_{12} \,,~ \\
\beta^\m_{xz}=\beta^\m_{zx}&= \frac{\gamma_{11}-\gamma_{22}}{2}    \,,~ \\
\beta^\m_{zz}&= \frac{\beta_{11}+\beta_{22} -2(\beta_{12}-\gamma_{12}) }{4}      \,.
}

\subsubsection*{Topological properties in two-dimensions}
The soft modulus vector field $\m$ \Eqref{Eq:App:M} can be associated with non-trivial 
topological properties, by considering the properties of the corresponding unit vector 
field $\n:=\m/\Norm{\m}$. Indeed, the unit vector $\n$ is a map from the one-point 
compactification of the plane ($\groupR{2}\cup\{\infty\}\cong \groupS{2} $) onto the 
two-sphere target space spanned by $\n$. That is $\n: \groupS{2}\mapsto \groupS{2}_\n$, 
which is classified by the homotopy class $\pi_2(\groupS{2}_\n)\in\groupZ{}$, 
thus defining the topological invariant, \ie the degree of the map, as   
\Equation{Eq:App:N:Charge}{
   \Q({\n})=\frac{1}{4\pi} \int_{\groupR{2}}
   \n\cdot\partial_x \n\times \partial_y \n\,\,
  d x d y \,.
}
Note that $\n$ is ill-defined when $\Norm{\m}=0$. On the other hand, whenever 
$\Norm{\m}\neq0$, the corresponding configuration have an integer topological charge 
$\Q(\n)\in\groupZ{}$. In a way, $\Q(\n)$ counts the number of times the pseudo-spin 
texture of $\n$ wraps the target two-sphere. The topological invariant 
\Eqref{Eq:App:N:Charge}, the index of the map $\n$, can be expressed directly in terms 
of the soft modulus vector field $\m$. This is easily done by replacing $\n$ with its 
actual definition $\n:=\m/\Norm{\m}$. The topological invariant thus reads as
\Equation{Eq:App:M:Charge}{
   \Q(\m)=\frac{1}{4\pi} \int_{\groupR{2}} \frac{
   \m\cdot\partial_x \m\times \partial_y \m}{\Norm{\m}^3}\,\,
  d x d y \,.
}
Here again, the integrand is obviously ill-defined when $\Norm{\m}=0$. Whenever 
$\Norm{\m}\neq0$, the corresponding configuration have an integer topological 
charge $\Q(\m)\in\groupZ{}$. The quantization of $\Q(\m)$ implies that the magnetic 
flux is quantized as well according to $\int \B_z = \Phi_0 \Q$, where $\Phi_0=-2\pi/e$ 
is the flux quantum.

It should be emphasized that unlike the flux quantization condition for the 
superconducting states, which is related to the $\groupU{1}$ topological invariant, 
the condition \Eqref{Eq:App:M:Charge} is also valid in the non-superconducting phase. 
Indeed, the quantization in the superconducting state is given by the $\groupU{1}$ 
topological invariant, which is related to the total phase winding at spatial infinity 
(the usual winding number). In the fermion quadrupling resistive state, the total phase 
of $\Psi$ is disordered and the $\groupU{1}$ invariant does not exist. On the other, 
since it is associated only with the relative phases, $\Q(\m)$ is the quantity that 
defines the flux quantization.

It is worth emphasizing that the topological charge \Eqref{Eq:App:N:Charge} is an 
integer, when integrated over the infinite plane $\groupR{2}$, or at least an large 
enough domain $\Omega\subset\groupR{2}$.

\subsection{Parameter sets}
\label{Sec:Parameters}

The essential features can be qualitatively summarized as follows: First, all the 
coefficients involving a $y$ index vanish: $\alpha^\m_y=\beta^\m_{ay}=\beta^\m_{ya}=0$. 
Moreover, the criterion for the condensation is
$\alpha^{\m\,2}_0>\alpha^{\m\,2}_x+\alpha^{\m\,2}_z$, and also 
$\beta^\m_{00},\beta^\m_{zz}>0$. 
The effect of the time-reversal symmetry operation for the soft modulus vector 
is a reflection of $\m$ on the $xz$-plane of the target space: 
\Equation{Eq:App:TRS}{
{\cal T}(\Psi)=\Psi^* ~~\Leftrightarrow~~
{\cal T}(\m)=(m_x,-m_y,m_z)	\,.
}

\newcolumntype{P}[1]{>{\centering\arraybackslash}p{#1}}
\newcommand{\COLSIZ}{0.0675}
\begin{table*}[!htb]
  \centering
 \begin{tabular} 
{|P{0.175\linewidth}|
P{0.075\linewidth}|P{\COLSIZ\linewidth}||
P{\COLSIZ\linewidth}|P{\COLSIZ\linewidth}|P{\COLSIZ\linewidth}||
P{0.075\linewidth}|P{\COLSIZ\linewidth}||
P{0.075\linewidth}|P{\COLSIZ\linewidth}||
}\hline 
	 	 Parameters of the	& $\alpha^\m_0$ 	& $\alpha^\m_x$ 	& $\alpha^\m_z$ 	
	 						& $\beta^\m_{00}$ 	& $\beta^\m_{0x}$	& $\beta^\m_{0z}$ 	
	 						& $\beta^\m_{xx}$	& $\beta^\m_{xz}$ 	& $\beta^\m_{zz}$	
							\\
 		 effective model	& $(\times10^{-1})$	& $(\times10^{-2})$	& $(\times10^{-2})$	
 							& $(\times10^{-1})$	& $(\times10^{-2})$	& $(\times10^{-2})$	
 							& $(\times10^{-1})$	& $(\times10^{-2})$	& $(\times10^{-1})$
 							\\
 \hline \hline %
 							& -8.2605			&  8.0923		& -6.3919	
 							&  4.5431			& -6.5085		&  1.4987	
 							&  2.1940			& -0.0875		&  2.3491
 							\\
 
 \hline  \hline   
  \end{tabular}
  \caption{
  Coefficients of the Ginzburg-Landau free energy functional 
  that correspond to the various numerical simulations reported in the main body 
  of the text. Starting from the microscopic model \Eqref{Eq:App:Model3BandB1},  
  the parameters of the coupling matrix are $u_{eh}=0.45$ and $u_{hh}=0.5$, and the 
  coefficients of the gradient term are $K^{(1)}=0.5$, $K^{(2)}=0.35$, $K^{(3)}=0.45$ 
  and the temperature parameter is $T/T_c=0.25$. Next, the coefficients of the 
  diagonalized Ginzburg-Landau model \Eqref{Eq:App:FreeEnergy:diagonal} are evaluated 
  using the formulas \Eqref{Eq:App:Coefficients:bilin}, \Eqref{Eq:App:Coefficients:quart:1}
  and \Eqref{Eq:App:Coefficients:quart:2}. 
  Finally the coefficients of the soft modulus effective model are obtained with the 
  formulas \Eqref{Eq:App:M:Coefficients:1}, \Eqref{Eq:App:M:Coefficients:2}, 
  and \Eqref{Eq:App:M:Coefficients:3}.
}
  \label{Tab:GL-coefficents}
\end{table*}

A typical value of the parameter set, when obtained from the microscopic model, 
is given in the Table \ref{Tab:GL-coefficents}.

\subsection{Modulation of the parameters}

Inhomogeneities in a sample typically result in spatially varying parameters of the 
Ginzburg-Landau model. In the system with broken time-reversal symmetry this can result 
in gradients of both densities and relative phases. This can in principle produce 
spontaneous magnetic fields. 
As emphasized in the main body, the material has slight inhomogeneity in the doping level, 
and this results in relatively small local modulation of the superconducting critical 
temperature. This can be accounted for by implementing spatial modulation of the 
prefactors of the quadratic terms of the Ginzburg-Landau theory. 
For example, the parameters of the quadratic term of the original Ginzburg-Landau 
theory \Eqref{Eq:App:FreeEnergy} formally depend on the temperature, 
$a_{ii}\equiv a_{ii}(\tau)=a^{0}_{ii}[\tau-1]$. Hence, it may be possible to model 
the effect of temperature inhomogeneities by requiring a spatial dependence of the 
parameters $a_{ii}\equiv a_{ii}(\tau,\x)=a^{0}_{ii}[\tau(\x)-1]$. Different areas of 
an inhomogeneous sample indeed can have different local critical temperatures.
Such a local modification of the parameter was demonstrated to be responsible for the 
existence of spontaneous magnetic fields, in different models with time-reversal symmetry 
breaking states. These include the responses to linear thermal gradients 
\cite{Supp:Silaev.Garaud.ea:15,Supp:Grinenko.Weston.ea:21}, hotspot created by a 
laser pulse \cite{Supp:Garaud.Silaev.ea:16} but also the effect of impurities 
\cite{Supp:Maiti.Sigrist.ea:15,Supp:Lin.Maiti.ea:16}, and other inhomogeneous arrays 
\cite{Supp:Garaud.Corticelli.ea:18a,Supp:Vadimov.Silaev:18}.

At the level of the effective model, implementing smoothly spatially varying amplitudes 
of the individual components,  can be modelled by small spatial variations of the 
coupling constants $\alpha^\m_0$ and $\alpha^\m_z$ accordingly, given the relations 
\Eqref{Eq:App:Coefficients:bilin} and \Eqref{Eq:App:M:Coefficients:1}.
In the main body, we considered random modulations in the form of 
$\tau(\x)=\tau_0[1+\delta\tau\,\mathrm{ran}(\x)]$ where $\mathrm{ran}(\x)$ is a smooth 
random surface. Here $\tau_0$ is the nominal reduced temperature, $\delta\tau$ is the 
amplitude of the thermal variation.
The idea to construct a random, smoothly varying quantity is to represent it as a Fourier 
series with random coefficients:
\Align{Eq:App:Random}{
f(\x)=-c_0 &+ \sum_{i=1}^{N_x}\sum_{j=1}^{N_y}c_{ij}
\cos 2\pi\left(\frac{ix}{L_x}+\frac{jy}{L_y}\right) \nonumber \\
+&s_{ij}\sin 2\pi\left(\frac{ix}{L_x}+\frac{jy}{L_y}\right) \,,
}
were $L_x$ and $L_y$ are the dimensions of the box that bounds the numerical domain.
$N_x$ and $N_y$ are cut-off in the Fourier expansion, and the coefficients $c_{ij}$ 
and $s_{ij}$ are random numbers $\in[-0.5:0.5]$.


\section{Numerical methods} 
\label{Sec:App:Numerics}

In the numerical investigations in the main body of the paper, we use Finite-Element 
Methods (FEM) (see e.g. \cite{Supp:Hutton,Supp:Reddy}) to handle the spatial discretization 
of the problem. In practice we use the finite-element framework provided by 
the FreeFEM library \cite{Supp:Hecht:12}. Within this finite-element framework, 
the minimization of the free energy is addressed using a non-linear conjugate gradient 
algorithm \cite{Supp:Fletcher.Reeves:64,Supp:Polak.Ribiere:69,Supp:Polyak:69,Supp:Shewchuk:94}.

\subsection{Finite-element formulation}

We consider the domain $\Omega$ which a bounded open subset of $\groupR{2}$ and denote 
$\partial\Omega$ its boundary. $H(\Omega)$ stands for the Hilbert space, such that a 
function belonging to $H(\Omega)$, and its weak derivatives have a finite $L^2$-norm. 
Furthermore. The Hilbert spaces of real-valued functions is equipped with the inner 
product $\langle\cdot,\cdot\rangle$, defined as:
\Equation{Eq:App:Inner}{
\ScalarProd{u}{v}=\int_\Omega uv  \,,~\text{for}~u,v\in H({\Omega}) \,.
}
The spatial domain $\Omega$ is discretized as a mesh of triangles using for the 
Delaunay-Voronoi algorithm, and the regular partition $\mathcal{T}_h$ of $\Omega$ refers 
to the family of the triangles that compose the mesh. Given a spatial discretization, 
the functions are approximated to belong to a \emph{finite-element space} whose properties 
correspond to the details of the Hilbert spaces to which the functions belong. We define 
${P}^{(2)}_h$ as the $2$-nd order Lagrange finite-element subspace of $H(\Omega)$. 
Now, the physical degrees of freedom can be discretized in their finite element 
subspaces. And we define the finite-element description of the degrees of freedom as 
$m_i\mapsto m_i^{(h)}\in \mathcal{P}^{(2)}_h$. This describes a linear vector space 
of finite dimension, for which a basis can be found. The canonical basis consists of 
the shape functions $\phi_k(\x)$, and thus 
\Equation{Eq:App:FEM:space}{
V_h(\mathcal{T}_h,\mathrm{P}^{(2)})=\Big\lbrace w(\x)=\sum_{k=1}^M w_k\phi_k(\x)
,\phi_k(\x)\in \mathrm{P}^{(2)}_h
\Big\rbrace\,.
}
Here $M$ is the dimension of $V_h$ (the number of vertices), the $w_k$ are called 
the degrees of freedom of $w$ and M the number of the degrees of freedom.
To summarize, a given function is approximated as its decomposition:
$w(\x)=\sum_{k=1}^M w_k\phi_k(\x)$, on a given basis of shape functions $\phi_k(\x)$ 
of the polynomial functions $\mathrm{P}^{(2)}$ for the triangle $T_{i_k}$. 
The finite element space $V_h(\mathcal{T}_h,\mathrm{P}^{(2)})$ hence denotes the 
space of continuous, piecewise quadratic functions of $x$, $y$ on each triangle of 
$\mathcal{T}_h$.

\begin{figure*}[!htb]
\hbox to \linewidth{ \hss
\includegraphics[width=0.95\linewidth]{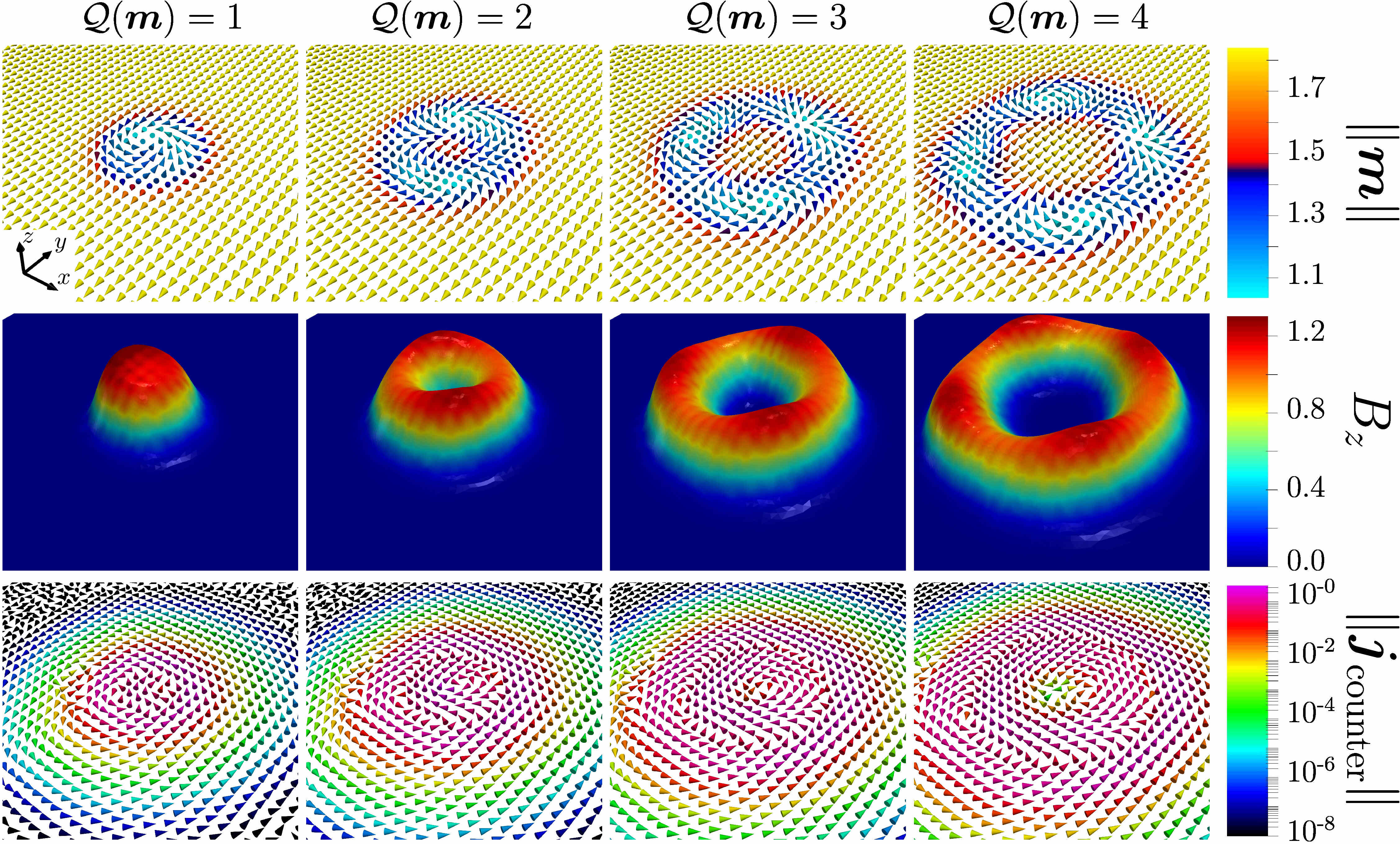}
\hss}
\vspace{1cm}
\hbox to \linewidth{ \hss
\includegraphics[width=0.95\linewidth]{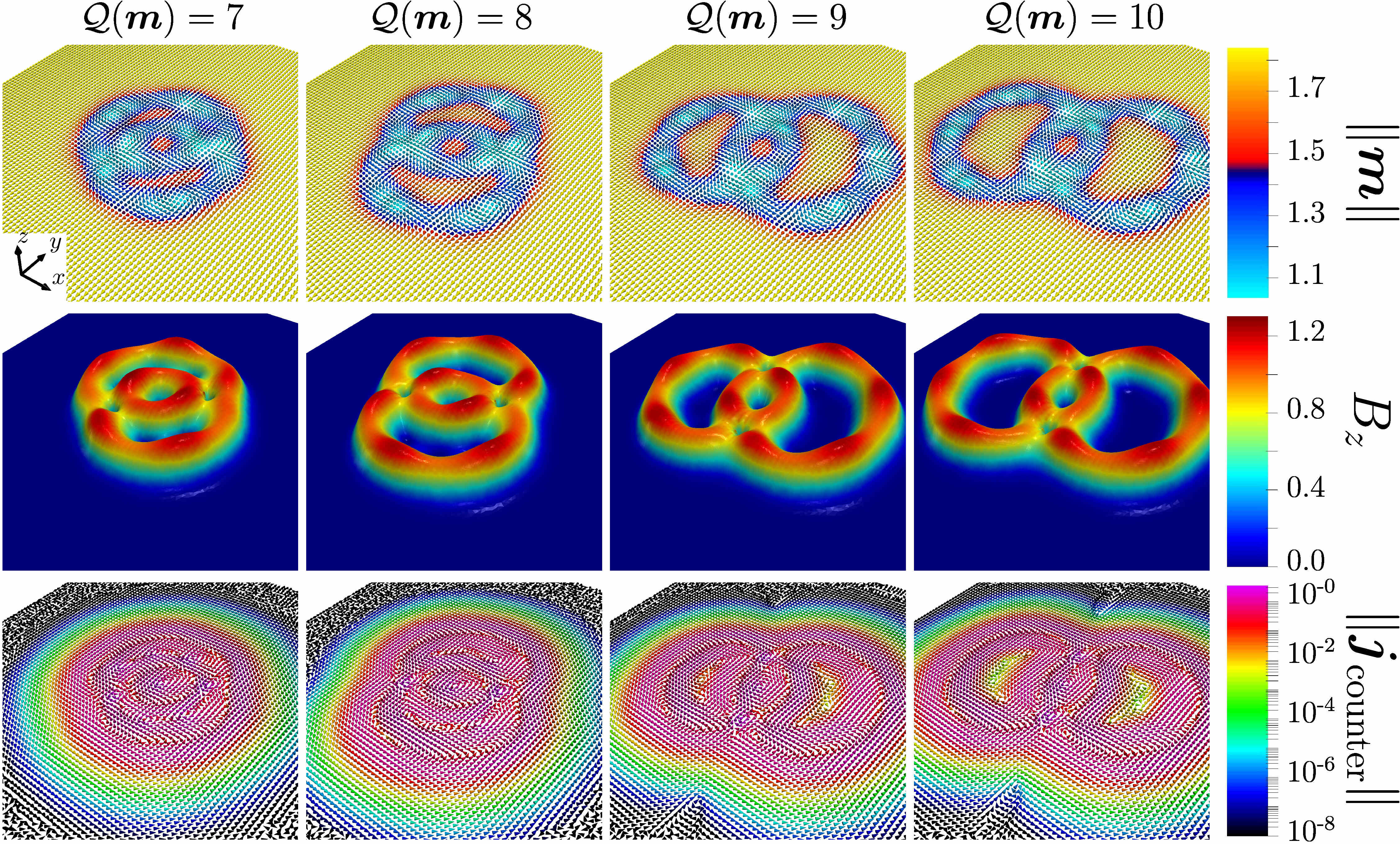}
\hss}
\vspace{0.1cm}
\caption{ 
Skyrmion solutions in a time-reversal symmetry broken state, for increasing values of the 
topological charge $\Q(\m)$. On each block, the panels on the top row display the texture of 
the four-fermion order parameter $\m$. The panels on the middle row show the associated 
magnetic field $\B$ \Eqref{Eq:App:M:Magnetic:1}, and the bottom row display the corresponding 
charge transferring counter-currents $\j_\text{counter}$ according to the Amp\`ere's law.
The parameters are the same as in the main body. Note that as can be seen from the densities 
of arrows, the configurations for $\Q(\m)$=7--10 are zoomed out, as compared to those for 
$\Q(\m)$=1--4.
}
\label{Fig:Skyrmions2}
\end{figure*}

\subsection{Initial guess: Skyrmions and domain-walls}

\begin{figure*}[!htb]
\hbox to \linewidth{ \hss
\includegraphics[width=0.95\linewidth]{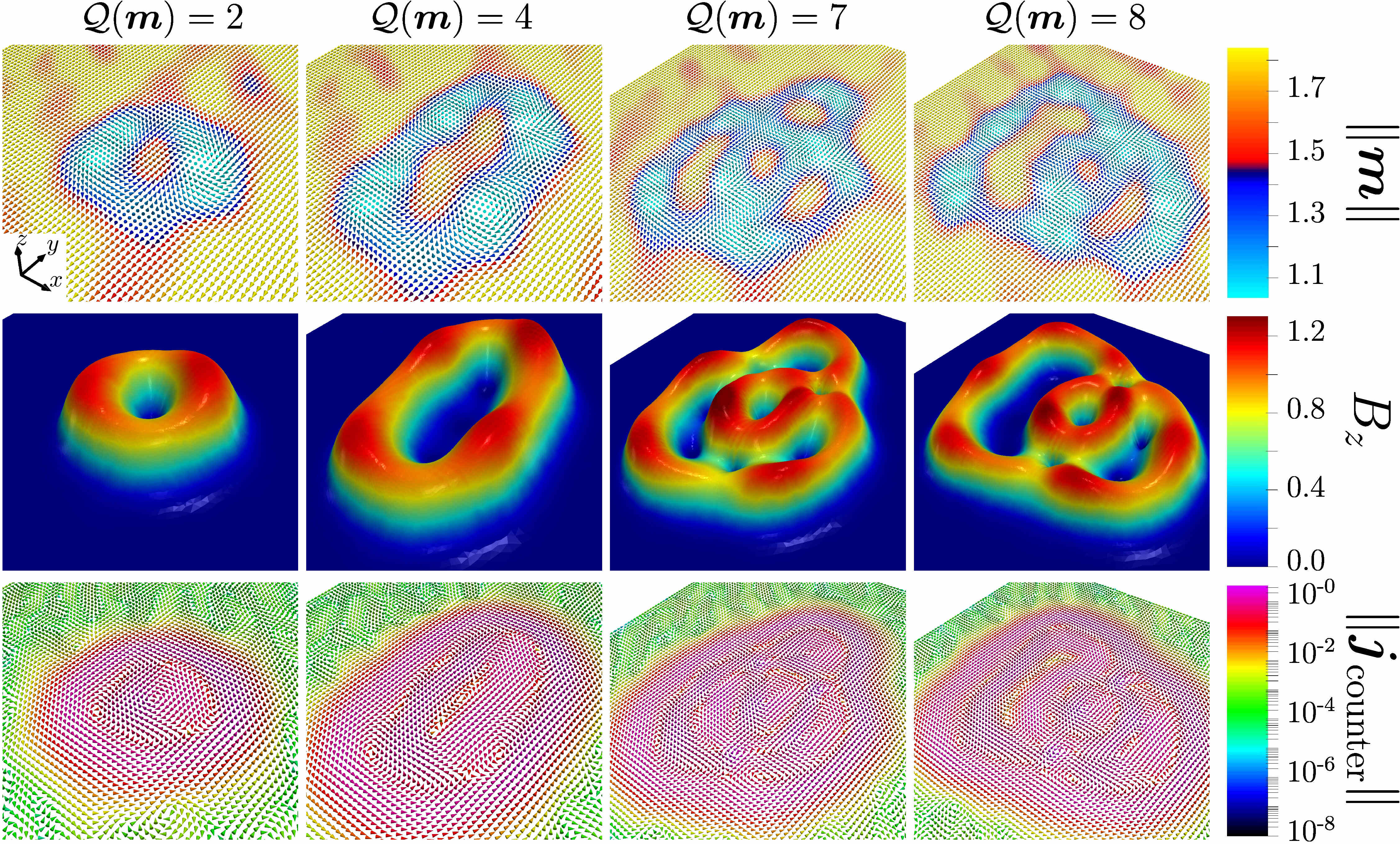}
\hss}
\vspace{0.25cm}
\caption{ 
Skyrmion solutions in the presence of inhomogeneities. On each block, the panels on the 
top row display the texture of the four-fermion order parameter $\m$. The panels on the 
middle row show the associated magnetic field $\B$ \Eqref{Eq:App:M:Magnetic:1}, and the 
bottom row display the corresponding charge transferring counter-currents $\j_\text{counter}$ 
according to the Amp\`ere's law.
The parameters are the same as in the main body. Note that as can be seen from the densities 
of arrows, the configurations for $\Q(\m)$=7,8 are zoomed out, as compared to those for 
$\Q(\m)$=2,4.
}
\label{Fig:Skyrmions3}
\end{figure*}

The skyrmions and the domain-walls are field configurations for the two-dimensional 
system. More precisely, either for the true two-dimensional system, or for a 
three-dimensional system with a translational invariance along the third direction $z$.
The initial guess is defined, such that the ground state would be $m_z=\pm1$. 
The configuration is then rotated using the rotation matrices $R_i$ and scaled to 
match the actual ground state $\hat{\m}$:
\Equation{Eq:App:IG:2d}{
\m = r_0 R_z(-\varphi_0) R_y(-\theta_0)  {\cal S}^{(sk)} \n^{(dw)}	\,.
} 
Here $r_0=\Norm{\hat{m}}$, $\theta_0=\arccos(\hat{m}_z/r_0)$, and 
$\phi_0 = \arctan(\hat{m}_y/\hat{m}_x)$ are the spherical coordinates of the ground 
state $\hat{\m}:=\mathrm{argmin}~V(\m)$, for the potential \Eqref{Eq:App:M:Potential}. 
Here $\n^{(dw)}$ is unit 3-vectors that encode the information about domain-walls 
\Eqref{Eq:App:DW}, and ${\cal S}^{(sk)}$ \Eqref{Eq:App:SK} is the function that 
imprints skyrmions on $\n^{(dw)}$.

The configuration that interpolates between the north and south pole of the unit 
sphere can be parametrized as follow: 
\SubAlign{Eq:App:DW}{
\n^{(dw)} &= \left(\sqrt{\frac{1-\Upsilon(\x)^2}{2}}, 
				   \sqrt{\frac{1-\Upsilon(\x)^2}{2}}, \Upsilon(\x)  \right) \,, \\
\Upsilon(\x) &= \tanh\left( \frac{\x_\perp-\x_0}{\xi_{dw}} \right) 
\,,
}
where $\xi_{dw}$ determines the width of the domain-wall. In \Eqref{Eq:App:DW}, 
$\x_0$ is the curvilinear abscissa that determines the position of the domain-wall, 
and $\x_\perp$ is the coordinate perpendicular to the domain-wall. In the absence 
of domain-walls, then $\n^{(dw)}=(0,0,1)$, simply points to the north pole.

The skyrmions are implemented by successively rotating the vector $\n^{(dw)}$. 
Namely, a set of $N_{sk}$ skyrmions is realized by successfully composing the 
rotations according to
\Equation{Eq:App:SK}{
{\cal S}^{(sk)} =  \prod^{N_{sk}}_{k=1} 
	R_z\left(\Phi_k(\x)\right) R_y\left(\Theta_k(\x)\right) \,.
}
Here again, $R_i$ are the rotation matrices, and the angles defining a given skyrmion 
are 
\SubAlign{Eq:App:Angles}{
\Phi_k(\x)   &= Q_k\arctan\left( \frac{y-y_k}{x-x_k} \right)		\,, \\
\Theta_k(\x) &= \pi \exp\left\{-\frac{(x-x_k)^2 + (y-y_k)^2}{2\xi_{sk}^2}\right\} \,.
}
The parameters $(x_k,y_k)$ determine the position of the core of the $k$-th skyrmion, 
of charge $Q_k$, and $\xi_{sk}$ determines the size of the skyrmions.

\section{Additional results}
\label{Sec:App:Additional}
\subsection{Skyrmions}

The model has a great variety of skyrmion solutions with different different 
topological charges. This can be see from \Figref{Fig:Skyrmions2} that displays 
several examples of stable skyrmions with topological charges $\Q(\m)$=1--10.

\subsection{Skyrmions on an inhomogeneous background}

The skyrmions displayed in the main body, as well as in \Figref{Fig:Skyrmions2}, 
are computed in the case of completely homogeneous parameters. However, as emphasized 
earlier, the materials can have slight inhomogeneities in doping level. This results 
in relatively small modulation of $T_c$, and also in modulation of relative densities and 
phases of the gaps. As emphasized in the main body, this yields spontaneous magnetic
fields. It is thus rather natural to question the effect that inhomogeneities can have 
on skyrmions. 
As already emphasized, in the considered model,  the skyrmions are fairly stable objects, 
we find that they survive in the presence of various kinds of inhomogeneities. 
This can be see from \Figref{Fig:Skyrmions3} that displays several examples of stable 
skyrmions (with topological charges $\Q(\m)$=2,4,7,8), in the presence of inhomogeneities. 
Clearly, the skyrmions are deformed by the inhomogeneities, yet, in this case they remain  
robust structures. See the conclusions of the main part of the paper regarding the 
reservations on stability of skyrmions beyond the effective model.

\subsection{Domain-walls}

\begin{figure}[!t]
\hbox to \linewidth{ \hss
\includegraphics[width=0.975\linewidth]{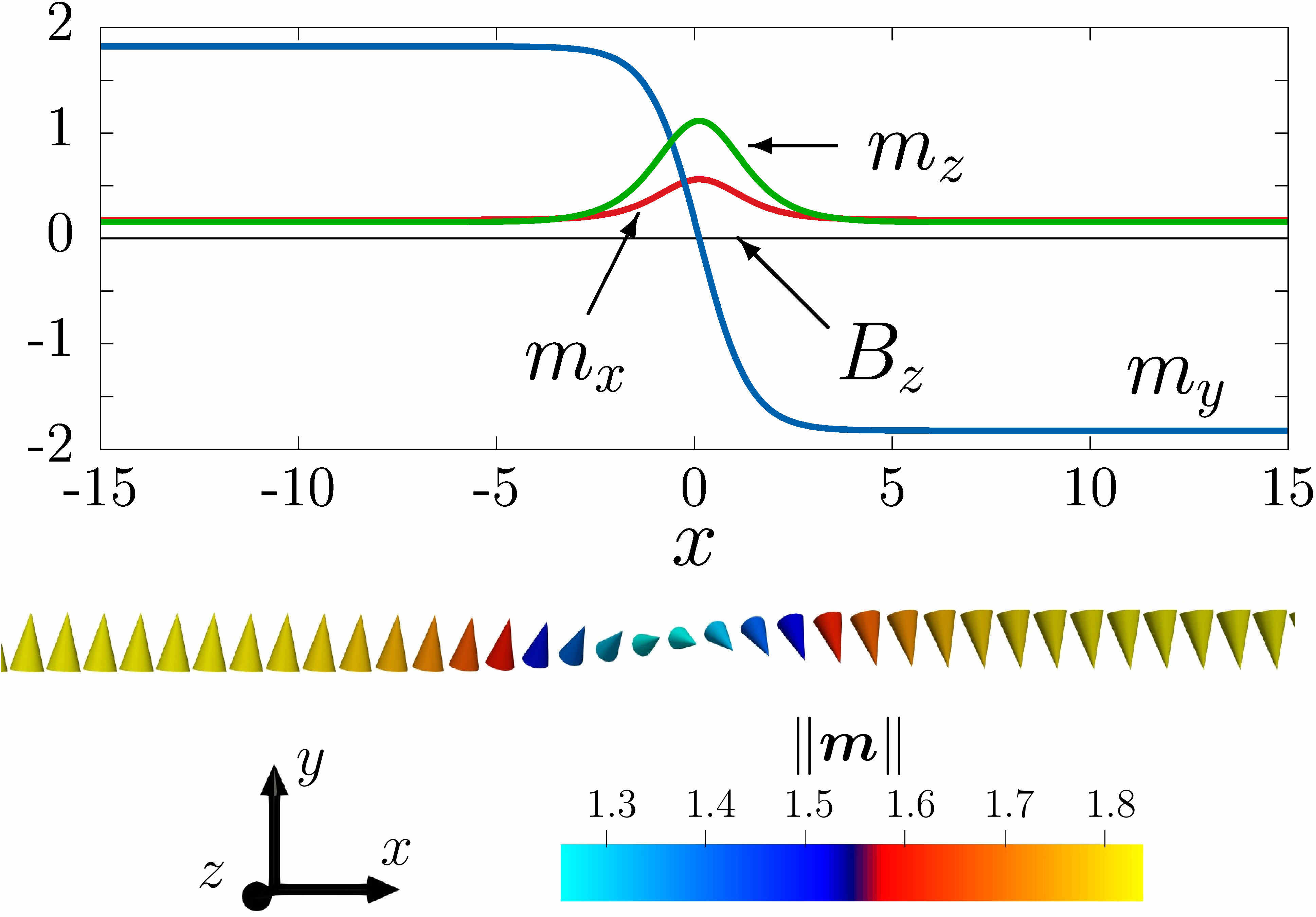}
\hss}
\caption{ 
A domain-wall that interpolates between the inequivalent time-reversal symmetry broken 
states. The top panel show the different components of the fermion quadrupling order 
parameter $\m$.The bottom panel shows the corresponding texture. The corresponding 
parameters are given in Sec.~\ref{Sec:Parameters}, and the gauge coupling $e=0.25$. 
The domain-walls are not associated with a magnetic field.
}
\label{Fig:DW}
\end{figure}

The fermion quadrupling resistive state, which precedes the $\sis$ state, spontaneously 
breaks the time-reversal symmetry. For the fermion quadrupling order parameter $\m$, 
the time-reversal operation \Eqref{Eq:App:TRS} implies that a state that breaks the 
time-reversal symmetry has $m_y\neq0$. In such a situation, there exist domain-wall 
excitations between both $s\pm is$ states, as illustrated in \Figref{Fig:DW}.

\end{document}